\documentclass[11pt]{article}
\usepackage{amssymb,amsmath,fullpage,times,epsf}
\usepackage{mathrsfs}
\usepackage{amssymb}
\usepackage{graphics}
\usepackage{graphicx}
\usepackage{subfigure}
\usepackage{float}
\usepackage{bm}
\usepackage{amsfonts}
\usepackage{color}

\allowdisplaybreaks

\newcommand{\me}{\mathrm{e}}
\newcommand{\mi}{\mathrm{i}}

\def\lam{\lambda}
\def\mb{\mathbf}
\def\ra{\rightarrow}
\def\Re{\mathrm{Re}}
\def\lim{\mathrm{lim}}

\def\PT{$\mathcal{P}\mathcal{T}$}

\def\eref#1{(\ref{#1})}

\usepackage[square,numbers,sort&compress]{natbib}
\def\Im{\mathrm{Im}}

\begin{document}

\date{}

\title{Rational Solitons in the Parity-Time-Symmetric Nonlocal Nonlinear Schr\"{o}dinger Model}

\author{Min Li$^{1}$, Tao Xu$^{2,}$\thanks{E-mail: xutao@cup.edu.cn}\,, and Dexin Meng$^{2}$
\\{\em 1 Department of Mathematics and Physics, }\\
{\em  North China Electric Power University, Beijing 102206, China} \\
{\em 2 College of Science, }\\
{\em China University of Petroleum, Beijing 102249, China} }\maketitle

\vspace{-5mm}

\begin{abstract}

In this paper, via the generalized Darboux transformation, rational soliton solutions are derived for the parity-time-symmetric nonlocal nonlinear Schr\"{o}dinger (NLS) model with the defocusing-type
nonlinearity. We find that the first-order solution can exhibit the elastic interactions of rational antidark-antidark, dark-antidark, and antidark-dark soliton pairs on a continuous wave background, but there is no phase shift for the interacting solitons. Also, we discuss the degenerate case in which only one rational dark or antidark soliton survives. Moreover, we reveal that the second-order rational solution displays the interactions between two solitons with combined-peak-valley structures in the near-field regions, but each interacting soliton  vanishes or evolves into a rational dark or antidark soliton as  $|z|\ra \infty$. In addition, we numerically examine the  stability of the first- and second-order rational soliton solutions.

$\;$



\end{abstract}

\newpage

\section{Introduction}

In classical quantum mechanics, a basic assumption is that the
Hamiltonian operator is Hermitian to ensure that every
physical observable is associated with a real
spectrum~\cite{Shankar}. In 1998, Bender and Boettcher proved that a
non-Hermitian Hamiltonian  also  has real and positive
eigenvalues  provided that it has the combined parity and time
reversal symmetry (usually called the $\mathcal{P}\mathcal{T}$
symmetry)~\cite{bender1}. Such pioneering work has led to the
complex extension of quantum mechanics~\cite{bender2}. In general, a
necessary condition for a Hamiltonian $H=\hat{p}^2/2+V(x)$ to be
$\mathcal{P}\mathcal{T}$-symmetric is that the complex potential
satisfies $V(x)=V^{*}(-x)$, where $\hat{p}$ denotes the momentum
operator~\cite{bender1,bender2,bender3}. Also, the notion of
$\mathcal{P}\mathcal{T}$ symmetry has  been applied to other areas
of theoretical physics, including Lie algebra~\cite{Lie}, complex
crystals~\cite{crystal}, quantum chromodynamics~\cite{Markum},
Bose--Einstein condensates~\cite{BEC}, classical
mechanics~\cite{mechanics}, and so forth.

Owing to the similarity between the paraxial equation of diffraction
in optics and the linear Schr\"{o}dinger equation in quantum
mechanics~\cite{OL2007,Makris1}, it is regarded that optics can
provide a fertile ground for realizing and testing the \PT-related
concepts~\cite{OL2007,Makris1,Makris2,Lin,Guo,Ruter}. An optical
potential respecting the $\mathcal {P}\mathcal {T}$ symmetry can be
realized in the complex refractive index distribution
$n_0+n_R(x)+\mi\, n_{I}(x)$, where the real index profile $n_R(x)$
should be an even function while the gain or loss component $n_I(x)$
must be odd, and $n_0$ is a constant background
index~\cite{OL2007,Makris1,Makris2}. It has been shown that
\PT-symmetric optical structures can exhibit  unique
characteristics such as double refraction, power oscillations,
spontaneous \PT~symmetry breaking, nonreciprocal diffraction
patterns, and unidirectional
invisibility~\cite{Makris1,Makris2,Lin,Guo,Ruter}. In experiments,
\PT~symmetry breaking within the realm of optics has  been
observed~\cite{Guo,Ruter}, which has stimulated the development of
\PT~optical materials and optical elements~\cite{Castaldi,Regensburger}.

In nonlinear optics,  the \PT~symmetry has received considerable
attention in the last few
years~\cite{Musslimani1,Gauss,Harmonic,RosenMorse,Gap,Defect,NonlinearLattice,MixLattice,Vector,Breather,Rogue,Lumer,Sukhorukov}.
Musslimani et al. first suggested the existence of optical
solitons with the presence of the Scarff II potential and periodic
\PT-symmetric potential~\cite{Musslimani1}. Later on, a lot of work
was devoted to the existence and stability of nonlinear
modes in different \PT-symmetric nonlinear systems, such as the
fundamental and higher-order solitons in the Gauss~\cite{Gauss},
harmonic~\cite{Harmonic}, and Rosen--Morse~\cite{RosenMorse}
\PT-symmetric potentials, gap solitons~\cite{Gap} and defect
solitons~\cite{Defect} in periodic \PT-symmetric potentials,
localized modes supported by \PT-symmetric nonlinear
lattices~\cite{NonlinearLattice}, lattice solitons in \PT-symmetric
mixed linear-nonlinear optical lattices~\cite{MixLattice}, and vector
solitons~\cite{Vector}, breathers~\cite{Breather} and rogue
waves~\cite{Rogue} in \PT-symmetric coupled waveguides.  Meanwhile,
researchers have studied the effects of nonlinearity on the
\PT~symmetry breaking~\cite{Lumer} and  dynamical characteristics of
a beam in \PT-symmetric optical nonlinear systems~\cite{Sukhorukov}.

Recently, Ablowitz and Musslimani proposed the following \PT-symmetric
nonlocal nonlinear Schr\"{o}dinger (NLS) equation~\cite{Ablowitz1}:
\begin{align}
\mi\,u_z(x,z) = u_{xx}(x,z) + 2\, \varepsilon\,u(x,z)u^*(-x,z)u(x,z)
\quad (\varepsilon= \pm 1)\,, \label{NNLS}
\end{align}
which is obtained from the standard NLS equation by replacing
$|u|^2u$ with $u(x,z)u^*(-x,z)u(x,z)$, where $u(x,z)$ denotes the
electric field envelope, $z$ is the spatial coordinate along the
propagation axis, $x$ is the transverse coordinate, $\varepsilon=\pm1$ denotes the focusing $(+)$ and
defocusing $(-)$ nonlinearity, and the star signifies the complex
conjugate. The \PT~symmetry of Eq.~\eref{NNLS} means
that the self-induced potential $V(x,z)=2\, \varepsilon\,u(x,z)u^*(-x,z)$ exactly
satisfies the relation $V(x,z)=V^{*}(-x,z)$, while the nonlocality
says that the value of the potential $V(x,z)$ at $x$
requires the information on $u(x,z)$ at $x$ as well as at
$-x$~\cite{Sarma}. 
Equation~\eref{NNLS} is integrable in the sense of
admitting the Lax pair and an infinite number of conserved
quantities. Therefore, its initial-value problem can be solved by the
inverse scattering transform (IST)~\cite{Ablowitz1}.  In addition,
the integrability of the discrete version of Eq.~\eref{NNLS} has
been established~\cite{Sarma,Ablowitz2}, and some other
\PT-symmetric nonlocal integrable models have also  been proposed~\cite{NLModels}.

In contrast with the standard NLS equation, the \PT-symmetric
nonlocal NLS model has many different properties. For the focusing
case, Eq.~\eref{NNLS} possesses both static bright and dark
solitons~\cite{Sarma,Khare2}, but its moving soliton  obtained by
the IST method contains a singularity~\cite{Ablowitz1}. In the
defocusing case, via the elementary Darboux transformation (DT), we
have revealed the nonsingular exponential soliton solutions on a
continuous wave (cw) background~\cite{LiXu}. Such exponential solitons in general appear in dark-dark,
antidark-antidark, antidark-dark or dark-antidark pairs, and can exhibit the usual elastic interactions.
It should be pointed out that the antidark soliton is another type of soliton existing in the normal dispersion
regime on a nonzero cw background~\cite{antidark}.
However, the  exponential solitons in Eq.~\eref{NNLS} will become unstable if the solution has a small shift  from the center of the \PT~symmetry, which has been confirmed by numerical simulation~\cite{Sarma,LiXu}.

We note that the kernel for constructing the elementary DT is made
up of linearly independent solutions of the Lax pair associated with
different spectral parameters~\cite{Matveev1}. Hence, the elementary
DT cannot deal with the degenerate cases when the spectral parameter
in the Lax pair reduces to some fixed value, so that the explicit
solutions are derived only in the exponential form (similar to the soliton
and breather solutions). In fact, via the generalized DT proposed by
Matveev~\cite{Matveev}, one can obtain the rational solutions in
such degenerate cases. In recent years, the generalized DT has been widely
used to construct the rational rogue wave solutions (which are
algebraically localized in any direction of the temporal-spatial
plane) of the NLS-type models~\cite{GDT,Roguewave}. In this paper,
we will construct the generalized DT of Eq.~\eref{NNLS} based on the
work in Ref.~\cite{LiXu}, and further reveal the rational soliton
phenomena on the cw background. In sharp contrast with the rogue wave
solutions, the rational solutions obtained in this work are
localized along the straight lines in the $xz$-plane, and can
display the profiles of the common exponential dark and antidark
solitons. Therefore, we call these two types of localized wave
structures as the rational dark (RD) and rational antidark (RAD)
solitons, respectively. Different from the exponential soliton
solutions obtained in Ref.~\cite{LiXu}, the first-order rational
soliton solution exhibits only the elastic interactions for the
RAD-RAD, RD-RAD, or RAD-RD soliton pairs, but the interacting solitons
do not experience the phase shift. We also find that the second-order rational solution displays the interactions between two solitons with combined-peak-valley structures in the near-field regions, but each interacting soliton will eventually vanish or evolve into a RD  or RAD soliton as  $|z|\ra \infty$.  Our numerical experiments show that the rational soliton solutions have good stability against small initial perturbations, but their stability will be destroyed   if the self-induced potential loses the \PT~symmetry  with respect to any point of $x$.

The structure of this paper is as follows: In
Sect.~\ref{Sec2}, we will construct the generalized DT based on
the elementary one in Ref.~\cite{LiXu}. In Sect.~\ref{Sec3}, we
will derive the first- and second-order rational soliton solutions of Eq.~\eref{NNLS} with $\varepsilon=-1$,
discuss the soliton interaction properties via asymptotic analysis,
and examine the stability of the rational soliton solutions by performing numerical experiments.
In Sect.~\ref{Sec4}, we will conclude this paper.

\section{Generalized Darboux Transformation}
\label{Sec2}

The Lax pair of Eq.~\eref{NNLS} can be written in the
form~\cite{Ablowitz1}
\begin{subequations}
\begin{align}
& \Psi_x=U\Psi=\begin{pmatrix}
\lam & u(x,z) \\
-\varepsilon u^*(-x,z) & -\lam
\end{pmatrix}\Psi , \label{CNLS4a} \\
& \Psi_z=V\Psi=\begin{pmatrix}
-2\mi\lam^2-\mi \varepsilon u(x,z) u^*(-x,z) & -2\,\mi \lam u(x,z) -\mi u_x(x,z)\\
2 \mi\varepsilon \lam   u^*(-x,z) - \mi \varepsilon u_x^*(-x,z) &
2\mi\lam^2 + \mi \varepsilon u(x,z) u^*(-x,z)
\end{pmatrix}\Psi,\label{CNLS4b}
\end{align}
\end{subequations}
where $ \Psi=(f, g)^{\rm{T}}$ (the superscript $\rm{T}$
represents the vector transpose) is the vector eigenfunction, $
\lambda $ is the spectral parameter, and Eq.~\eref{NNLS} can be
recovered from the compatibility condition $U_z - V_x + U\,V -
V\,U=0$.

On the basis of work described in Ref.~\cite{LiXu}, the  $N$th
 iterated elementary DT for Eq.~\eref{NNLS} can be
constituted by the eigenfunction transformation
\begin{align}
& \Psi[N]=T[N]\Psi, \quad   T[N]=
\begin{pmatrix}
\lam^N  - \sum\limits_{n=1}^{N}a_n(x,z) \lam^{n-1}  & - \sum\limits_{n=1}^{N}b_n(x,z) (-\lam)^{n-1}  \\
- \sum\limits_{n=1}^{N}c_n(x,z) \lam^{n-1}  & \lam^N  -
\sum\limits_{n=1}^{N}d_n(x,z)
(-\lam)^{n-1}   \\
\end{pmatrix}  \label{eigenfunctionTran}
\end{align}
and the potential transformation
\begin{align}
&  u[N](x,z) = u(x,z) +2\,(-1)^{N-1} b_{N}, \quad u^*[N](-x,z)=
u^*(-x,z) + 2\,\varepsilon c_{N} , \label{PotentialTrana}
\end{align}
where $ N $ represents the iterated time. The new eigenfunction
$\Psi[N]$ is required to satisfy the Lax pair in Eqs.~\eref{CNLS4a}
and~\eref{CNLS4b} with $u[N](x,z)$ and $u^*[N](-x,z)$ instead of
$u(x,z)$ and $u^*(-x,z)$, respectively. The functions $ a_n(x,z),
b_n(x,z), c_n(x,z)$, and $d_n(x,z)$ $( 1\leq  n \leq N )$ can be
determined from
\begin{align}
& T[N]\mid_{\lam=\lam_k}\Psi_k=\mb{0}, \quad
T[N]\mid_{\lam=\lam^*_k} \bar\Psi_k=\mb{0} \quad (1 \leq k \leq N),
\label{UndetCoeff}
\end{align}
where $\Psi_k=\big[f_k(x,z), g_k(x,z)]^{\rm{T}}$ and
$\bar{\Psi}_k=\big[g^*_k(-x,z), \varepsilon
f^*_k(-x,z)\big]^{\rm{T}}$ are the solutions of the Lax
pair in Eqs.~\eref{CNLS4a} and~\eref{CNLS4b} with $\lam = \lam_k$ and $\lam
= \lam^*_k$, respectively. In particular, via Cramer's rule, the functions $
b_{N}(x,z)$ and  $c_{N}(x,z) $ can be obtained in the determinant
form
\begin{align}
& b_N= (-1)^{N-1}\frac{\tau_{N+1,N-1}}{\tau_{N,N}},\quad c_N =
 \frac{\tau_{N-1,N+1}}{\tau_{N,N}},
\label{PotentialTranb}
\end{align}
with \begin{align}
\tau_{N, N}=
\begin{vmatrix}
F_{N\times N} & G_{N\times N} \\
\varepsilon \bar{G}_{N\times N} & \bar{F}_{N\times N}
\end{vmatrix},
\end{align}
where the block matrices $ F_{N\times N} =
\big[\lam_k^{m-1}f_k(x,z)\big]_{1 \leqslant k,m \leqslant  N}$,
$G_{N\times N}= \big[(-\lam_k)^{m-1}g_k(x,z)\big]_{1 \leqslant k,m
\leqslant  N}$,  $\bar{F}_{N\times N} =
\big[(-\lam^*_k)^{m-1}f^*_k(-x,z)\big]_{1 \leqslant k,m \leqslant
N}$, and $\bar{G}_{N\times N} =
\big[\lam^{*m-1}_k g^*_k(-x,z)\big]_{1 \leqslant k,m \leqslant
N}$.


Note that the elementary DT does not apply to the degenerate case
when $\lam_{k_1+1}, \dots, \lam_{k_2-1}\ra \lam_{k_1}$,
$\lam_{k_2+1}, \dots , \lam_{k_3-1} \ra \lam_{k_2}$, $\cdots$,
$\lam_{k_n+1}, \dots, \lam_N \ra \lam_{k_n}$, where $1=k_1 < k_2 <
\dots < k_n \leq N $, $1\leq n \leq N$, and $\lam_{k_i} \neq
\lam_{k_j} $ ($1\leq i <j \leq n$).
For such a degenerate case, the functions $a_n(x,z)$, $b_n(x,z)$,
$c_n(x,z)$, and $d_n(x,z)$ ($1\leq n\leq N$) in the Darboux matrix
$T[N]$ cannot be uniquely determined because the coefficient matrix
in Eq.~\eref{UndetCoeff} is singular. To overcome this problem, we
define $m_i= k_{i+1}-k_i-1$ for $1\leq i \leq n-1 $ and
$m_n=N-k_n$, and assume that $ f_{k_i+h} (x,z)=
f_{k_i}(x,z,\lam_{k_i+h} )$ and $g_{k_i+h}(x,z) =
g_{k_i}(x,z,\lam_{k_i+h} )$ if $m_i>0$, where $\lam_{k_i+h} =
\lam_{k_i} + \epsilon_i $,  $ 1\leq h \leq m_i$,
$\epsilon_i$ 
are small parameters, and $[f_{k_i}(x,z), g_{k_i}(x,z)]^{\rm{T}} $
corresponds to the solution of the Lax pair in Eqs.~\eref{CNLS4a}
and~\eref{CNLS4b} with $\lam=\lam_{k_i}$ ($1\leq i \leq n $).


Via the idea of Matveev's generalized DT~\cite{Matveev,GDT}, we
expand the elements in the block matrices $F_{N \times N}$, $G_{N
\times N}$, $\bar{F}_{N\times N}$, and $\bar{G}_{N\times N}$  in the
Taylor series form
\begin{subequations}\label{FGH2}
\begin{align}
& \lam_{k_i+h} ^{m-1} f_{k_i+h} (x,z,\lam_{k_i+h})
=\sum_{j=0}^{\infty}f_{k_i}^{(m-1,j)}(x,z)\epsilon_i^{j} ,
\label{EP1}
\\
& (-\lam_{k_i+h}) ^{m-1} g_{k_i+h}(x,z,\lam_{k_i+h})
=\sum_{j=0}^{\infty}g_{k_i}^{(m-1,j)}(x,z)\epsilon_i^{j},
\label{EP2}
\end{align}
\end{subequations}
with
\begin{align}
& f_{k_i}^{(m-1,j)}(x,z) =\frac{1}{j !}\frac{\partial^j
[\lam_{k_i+h}^{m-1} f_{k_i} (x,z, \lam_{k_i+h})]}{\partial
\lam^j_{k_i+h}}\Big|_{\epsilon_i=0}, \notag \\
& g_{k_i}^{(m-1,j)}(x,z) =\frac{1}{j !}\frac{\partial^j
[(-\lam_{k_i+h})^{m-1} g_{k_i} (x,z, \lam_{k_i+h})]}{\partial
\lam_{k_i+h}^j}\Big|_{\epsilon_i=0}, \notag
\end{align}
where $1\leq i \leq n $, $1\leq m\leq N$, $1\leq h \leq m_i$, and
particulary $f_{k_i}^{(0,0)}(x,z)=f_{k_i}(x,z)$,
$g_{k_i}^{(0,0)}(x,z)=g_{k_i}(x,z)$.

Taking the limit $\epsilon_i \ra 0$, one can find that the
coefficient matrix in Eq.~\eref{UndetCoeff} is no longer singular,
which means that all the undetermined functions in $T[N]$ can be
uniquely solved. Thus, the new potential transformations are given
as
\begin{align}
u[N](x,z) =u(x,z)+2\,\frac{\tau'_{N+1,N-1}}{\tau'_{N,N}}, \quad
u^*[N](-x,z)= u^*(-x,z) +2\,\varepsilon
\frac{\tau'_{N-1,N+1}}{\tau'_{N,N}}, \label{NPT}
\end{align}
with \begin{align}
\tau'_{N, N}= \begin{vmatrix}
 F'_{N \times N} & G'_{N \times N} \\
\varepsilon \bar{G}'_{N \times N} & \bar{F}'_{N \times N}
\end{vmatrix},
\end{align}
where
\begin{align}
F'_{N\times N}= \begin{pmatrix}
F'_1\\
\vdots\\
F'_n
\end{pmatrix}, \quad
G'_{N\times N}=\begin{pmatrix}
G'_1\\
\vdots\\
G'_n
\end{pmatrix}, \quad
\bar{F}'_{N\times N}= \begin{pmatrix}
\bar{F}'_1\\
\vdots\\
\bar{F}'_n
\end{pmatrix}, \quad
\bar{G}'_{N\times N}=\begin{pmatrix}
\bar{G}'_1\\
\vdots\\
\bar{G}'_n
\end{pmatrix},
\end{align}
and $F'_i$, $G'_i$, $\bar{F}'_i$, and $\bar{G}'_i$  ($1\leq i \leq
n$) are the block matrices $ F'_i=\left[
f_{k_i}^{(p,q)}(x,z)\right]_{0 \leq p \leq N-1 \atop 0 \leq q \leq
m_i}$, $G'_i=\left[ g_{k_i}^{(p,q)}(x,z)\right]_{0 \leq p \leq N-1
\atop 0 \leq q \leq m_i}$, $ \bar{F}'_i=\left[(-1)^p
f_{k_i}^{*(p,q)}(-x,z)\right]_{0 \leq p \leq N-1 \atop 0 \leq q \leq
m_i}$, and $\bar{G}'_i=\left[(-1)^p g_{k_i}^{*(p,q)}(-x,z)\right]_{0
\leq p \leq N-1 \atop 0 \leq q \leq m_i}$.

\vspace{3mm}

Therefore, the eigenfunction transformation in Eq.~\eref{eigenfunctionTran}
and potential transformation in Eq.~\eref{NPT} constitute the generalized
DT for Eq.~\eref{NNLS} when some of the spectral
parameters $\{\lam_k\}_{k=1}^N$ coincide with each other. It is
obvious that the elementary transformation in Eq.~\eref{PotentialTrana}
corresponds to the particular case of the generalized one in Eq.~\eref{NPT}
when $n=N$. To avoid the triviality of the DT, we require that $\lam_k$ cannot be a real number.  In the next section, we will use the generalized DT to construct the
rational soliton solutions of Eq.~\eref{NNLS} with $\varepsilon=-1$ on a cw background.

\section{Rational Solitons on a cw Background}
\label{Sec3}

It is easy to see that Eq.~\eref{NNLS} with $\varepsilon=-1$ has the plane wave
solution~\cite{LiXu} 
\begin{align}
u=\rho\,\me^{K\,x+\mi\,\Omega\, z +\mi \phi} \quad
(\Omega=2\,\rho^2-K^2),\label{PlanSol}
\end{align}
where $K$,  $\rho$, and $\phi $ are three real parameters. Obviously, such a plane wave solution is unstable
when $K\neq 0$. Thus, we take $K=0$ in Eq.~\eref{PlanSol} and substitute it into the Lax pair in Eq.~\eref{CNLS4a}
and~\eref{CNLS4b} with $\lam=\lam_k$ ($1 \leq k \leq N$), which gives the solution for
$f_k$ and $g_k$ as follows:
\begin{align}\label{laxpairsolution}
\begin{pmatrix}
f_k\\
g_k
\end{pmatrix}=
\begin{pmatrix}
\me^{\frac{ 2\,\mi  \rho^2 z +\mi \phi}{2}}\big(\alpha_k \me^{ \mu_k  \chi_k}+\beta_k \me^{ -\mu_k  \chi_k }\big)\\
\me^{-\frac{ 2\,\mi  \rho^2 z +\mi \phi}{2}}\big(\frac{(\mu_k-
\lam_k) \alpha_k }{\rho} \me^{\mu_k \chi_k } - \frac{(\mu_k + \lam_k
) \beta_k}{\rho} \me^{ -\mu_k \chi_k }\big)
\end{pmatrix},
\end{align}
with $\mu_k =\sqrt{\lam_k^2 + \rho^2}$ and $\chi_k =x- 2\,\mi\, \lam_k
z$, where $ \alpha_k$ and $\beta_k $ ($1 \leq k \leq N$) are free
complex parameters.  It should be noted that the solution in Eq.~\eref{laxpairsolution}
will reduce to a rational one if $\lam_k= \mi \sigma \rho $.
Hence, in order to derive the rational solutions, we take  $ \lam_1= \mi \sigma \rho $ and
let $\lam_2, \dots, \lam_N $ all degenerate to $ \lam_1$, which corresponds to
$n=1$, $k_1=1$, and $m_1=N-1$. Following the expansions in Eqs.~\eref{EP1} and~\eref{EP2},
we can obtain the
formulas for $f_1^{(m-1,j)}(x,z)$ and $g_1^{(m-1,j)}(x,z)$  ($1\leq
m\leq N+1$; $j=0,1,2, \dots $). For simplicity, in the following
calculations, we set $\epsilon_i=\delta^2$,
$\alpha_k=\me^{\mu_k\sum_{j=1}^{\infty}s_j \delta^{2(j-1)} }$,
$\beta_k=-\me^{-\mu_k\sum_{j=1}^{\infty}s_j \delta^{2(j-1)} }$ for
$1\leq k \leq N $, where $s_j$ are arbitrary complex numbers.

\vspace{5mm}

\noindent{\em 3.1.  First-order rational soliton solution}

By truncating the expansions in Eqs.~\eref{EP1} and~\eref{EP2} at
$j=1$, we obtain the first-order rational soliton solution as
follows:
\begin{align}
u_1 = \rho\,\me^{2 \mi \rho^2z +\mi \phi }\bigg[ 1  - \frac{(2 \rho
\xi+K-\mi \sigma ) \left(2 \rho \eta -K^* + \mi \sigma\right)}{2
\rho^2 \xi \eta +\rho K  \eta - \rho K^* \xi   - \frac{1}{2}
\left(|K|^2 + 1\right)} \bigg], \label{soliton}
\end{align}
where $\xi =x+2 \,\sigma \rho z $,  $\eta= x - 2\,\sigma \rho z $,
$K= 2 \rho s_1 + \mi \sigma $, and  $\sigma=\pm 1$. It can be proved
that the solution in Eq.~\eref{soliton} has no singularity if and only if the
parameter $s_1$ satisfies the condition
\begin{align}
\Im(s_1) \neq -\frac{ \sigma}{2\,\rho }. \label{condition}
\end{align}
Under this condition, we perform an asymptotic analysis of
the solution in Eq.~\eref{soliton} so as to clarify the dynamical behavior
underlying  the solution.
\begin{figure}[H]
 \centering
\subfigure[]{\label{Fig1a}
\includegraphics[width=2in]{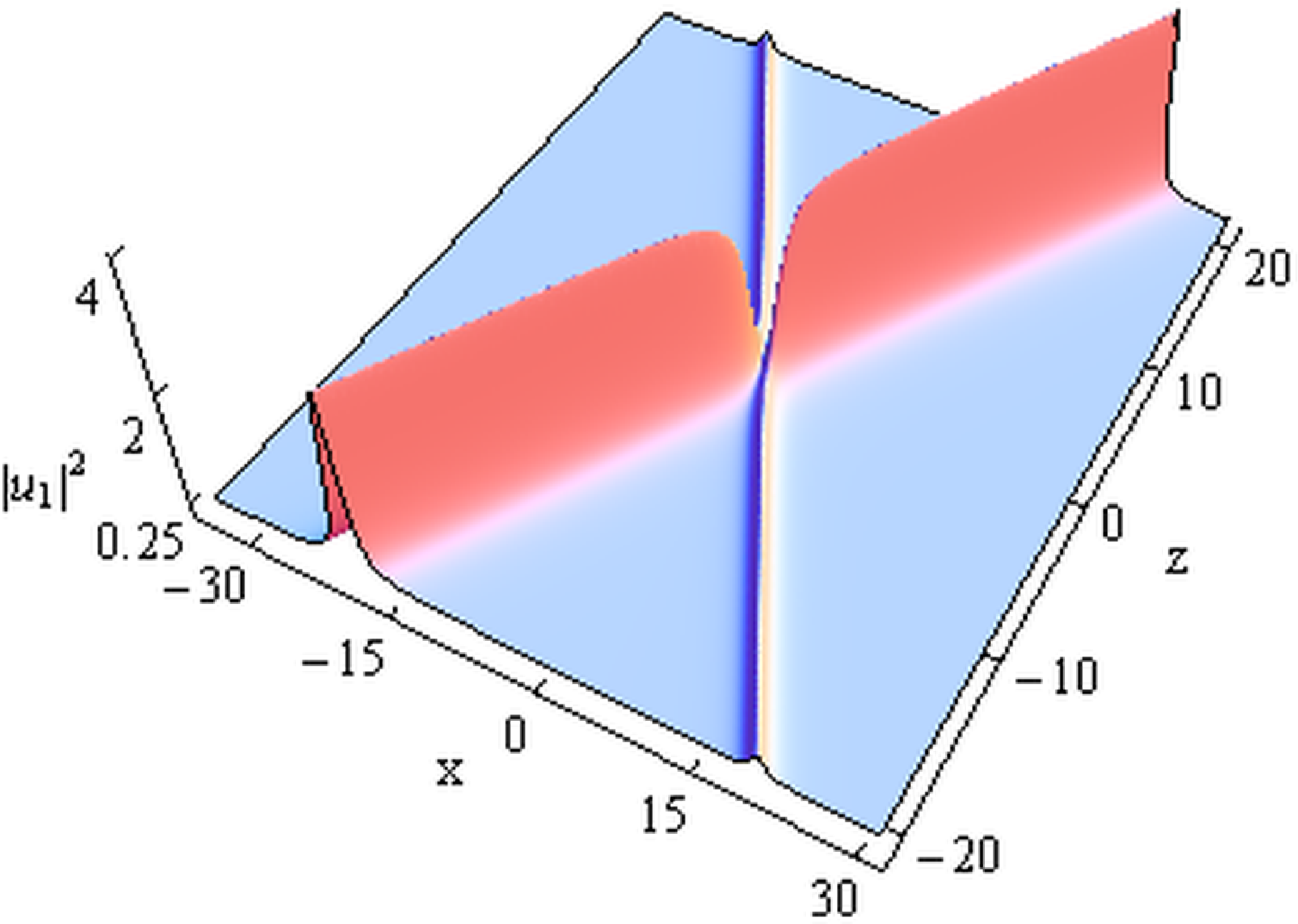}}\hfill
\subfigure[]{ \label{Fig1b}
\includegraphics[width=2in]{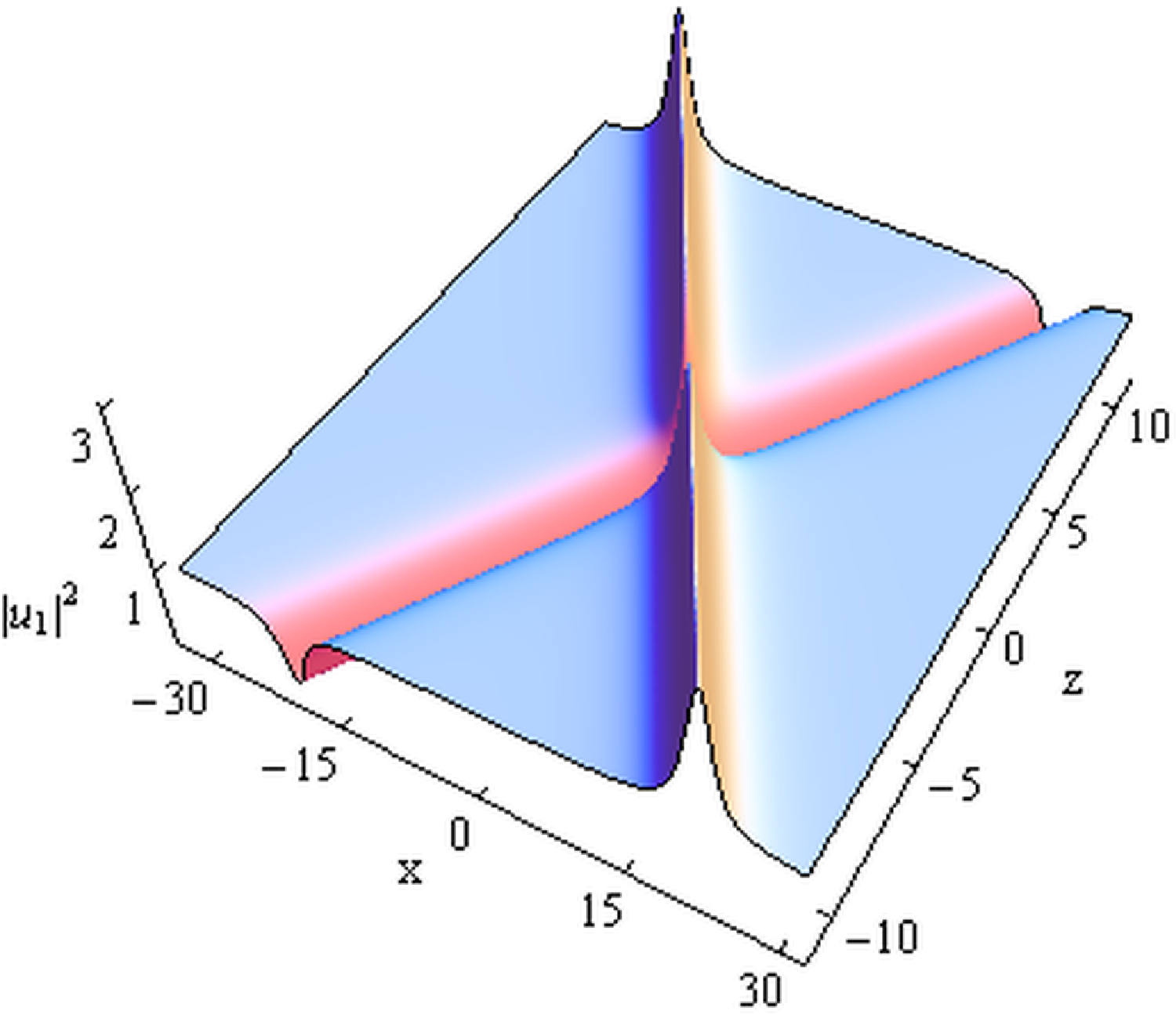}}\hfill
\subfigure[]{ \label{Fig1c}
 \includegraphics[width=2in]{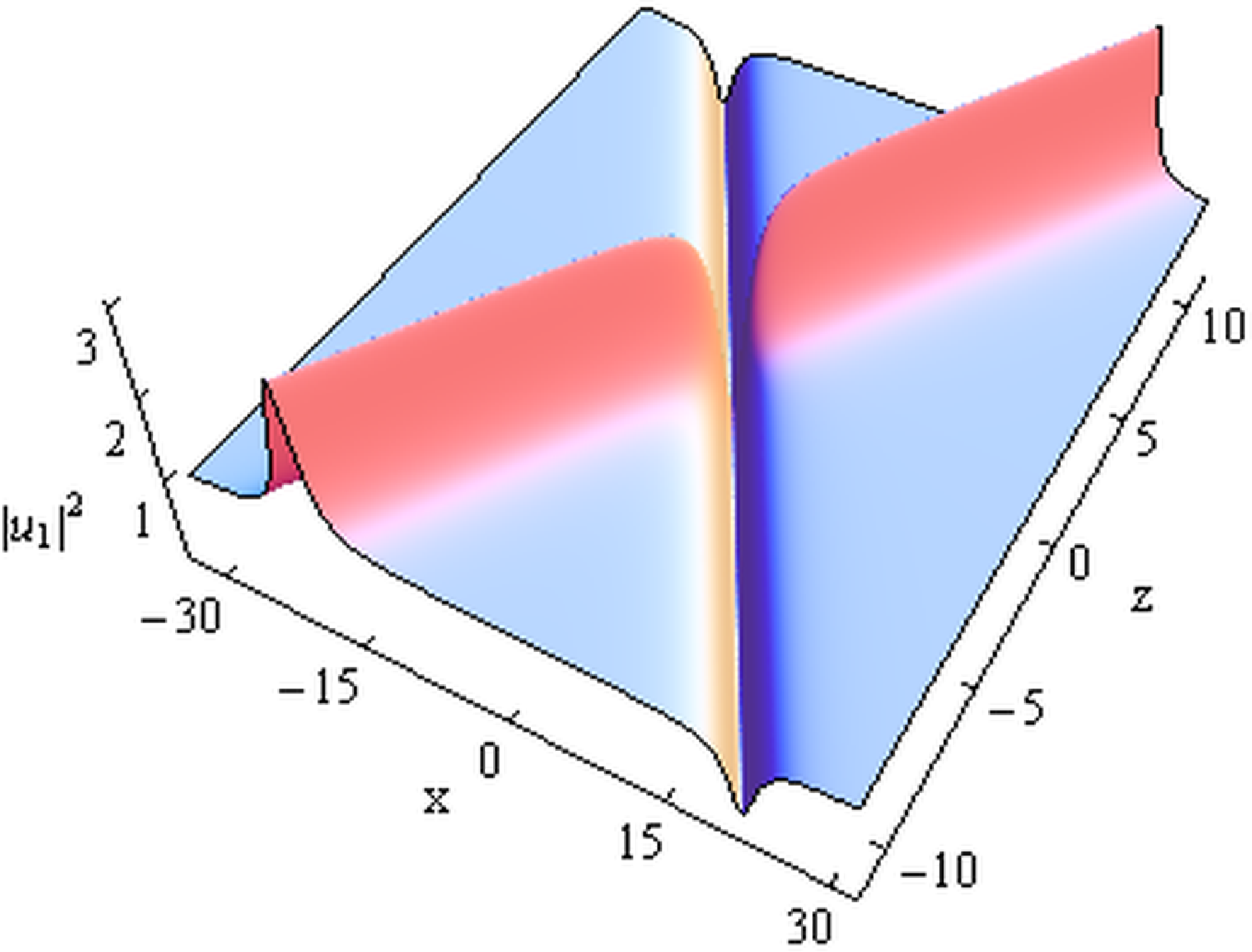}}
\caption{\small Three types of elastic two-soliton interactions via
the solution in Eq.~\eref{soliton}:  (a) RAD-RAD interaction with $\rho=0.5$,
$\phi=0$, $\sigma=1$, $s_1=1-0.2\,\mi$. (b)  RAD-RD  interaction
with $\rho=1$, $\phi=0$, $\sigma=1$, $s_1=1-2\,\mi$. (c) RD-RAD
interaction with $\rho=1$, $\phi=0$, $\sigma=1$,
$s_1=1+\mi$.\label{Fig1} }
\end{figure}
First, we obtain the asymptotic expression of
the solution in Eq.~\eref{soliton} along the line $ x + 2\,\sigma \rho z \sim
0$ as $|z| \ra \infty$ as follows:
\begin{subequations}
\begin{align}
& u_1 \ra u^{\rm{I}}_1:= \rho\,\me^{2\,\mi \rho^2 z +\mi
\phi}\bigg[1- \frac{4 \rho (x+2 \,\sigma \rho z) + 4 \rho \,s_1}{2 \rho (x+2 \,\sigma \rho z) + K}\bigg], \label{asymp1a}  \\
& |u_1|^2 \ra |u^{\rm{I}}_1|^2 = \rho^2 \bigg[1  -\frac{8\, \sigma
\rho \Im(s_1)}{|2  \rho (x+2 \,\sigma \rho z) + K|^2 }\bigg].
\label{asymp1b}
\end{align}
\end{subequations}
For the cases $ \sigma \Im(s_1)> 0 $ and $ \sigma
\Im(s_1)< 0 $, the intensity $|u^{\rm{I}}_1|^2$  can respectively exhibit the RD soliton beneath the cw background
$u=\rho\,\me^{2\,\mi \rho^2 z +\mi \phi}$ and the RAD soliton on top
of the same background, and  the valley and peak are both localized along
the line $x + 2\,\sigma \rho z  + \Re(s_1) =0 $.   Along the
line $x + 2\,\sigma \rho z  + \Re(s_1) =0 $,
$|u^{\rm{I}}_1|^2$ reaches the minimum for $ \sigma \Im(s_1)>
0 $ and the maximum for $ \sigma \Im(s_1)< 0 $. Thus, the
height of the RAD soliton or the depth of  the RD soliton is obtained as $A_1 = \frac{8 \rho^3 |
\Im(s_1)|}{\left[\sigma +2 \rho \Im(s_1)\right]^2}$.
The velocities of both the RD and RAD  solitons are proportional to
the amplitude of the cw wave, that is,  $v_1 = -2\,\sigma \rho $.
In particular, for $ \sigma
\Im(s_1)> 0 $, we use $\omega_1= \arccos \big[\frac{8 \rho |
\Im(s_1)|}{\left(2 \rho |\Im(s_1)| +1\right){}^2}\big]$
to characterize the darkness of the RD soliton. If $ \omega_1
=\frac{\pi}{2}$ (i.e., $|\Im(s_1)| = \frac{1}{2\,\rho}$),
$u^{\rm{I}}_1$ represents a rational black soliton; while for $
0< \omega_1 < \frac{\pi}{2} $ (i.e., $|\Im(s_1)| \neq 0,
\frac{1}{2\,\rho}$), it describes a rational gray soliton.

\begin{figure}[H]
 \centering
\subfigure[]{\label{Fig2a}
\includegraphics[width=2.5in]{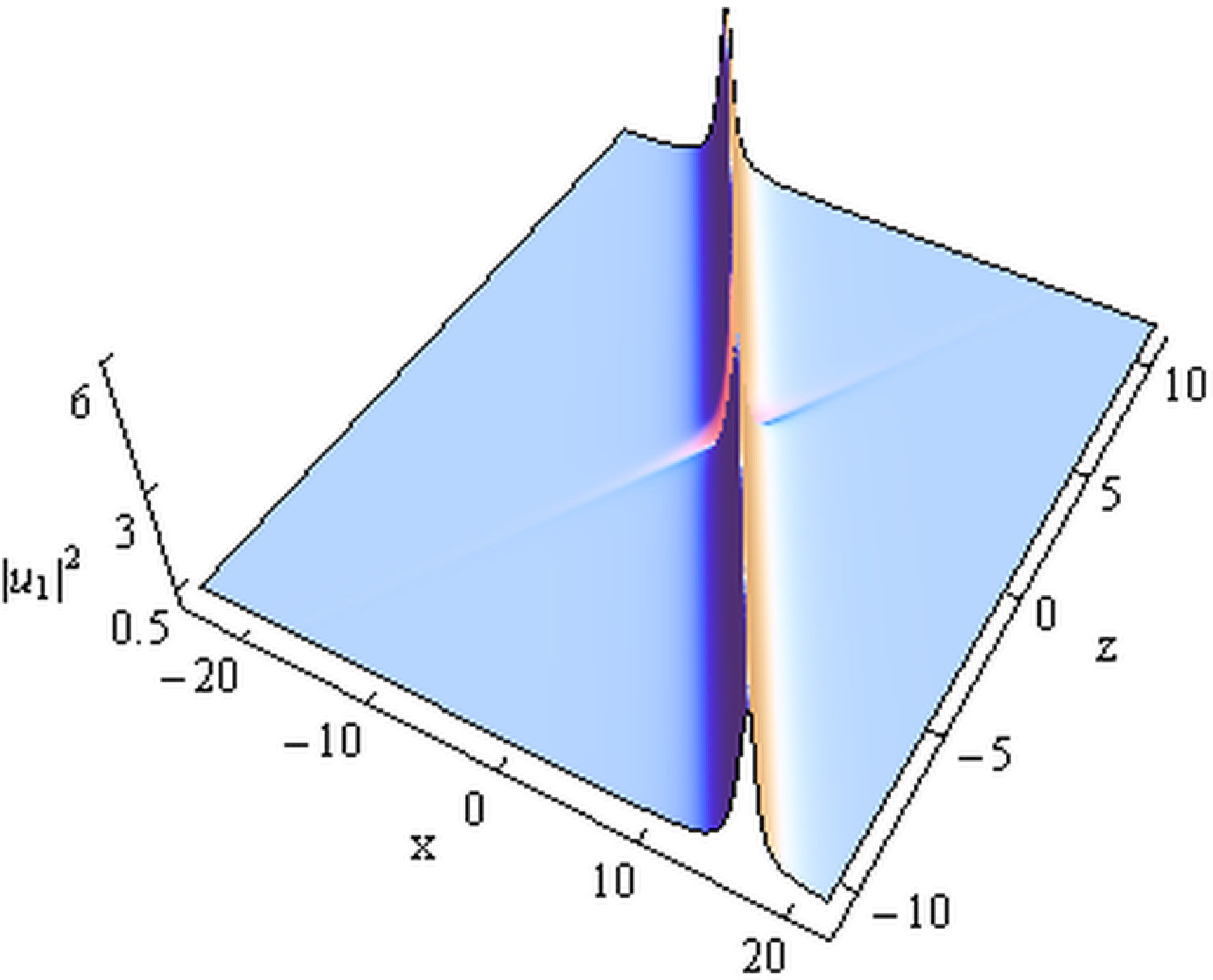}}\hfill
\subfigure[]{ \label{Fig2b}
\includegraphics[width=2.5in]{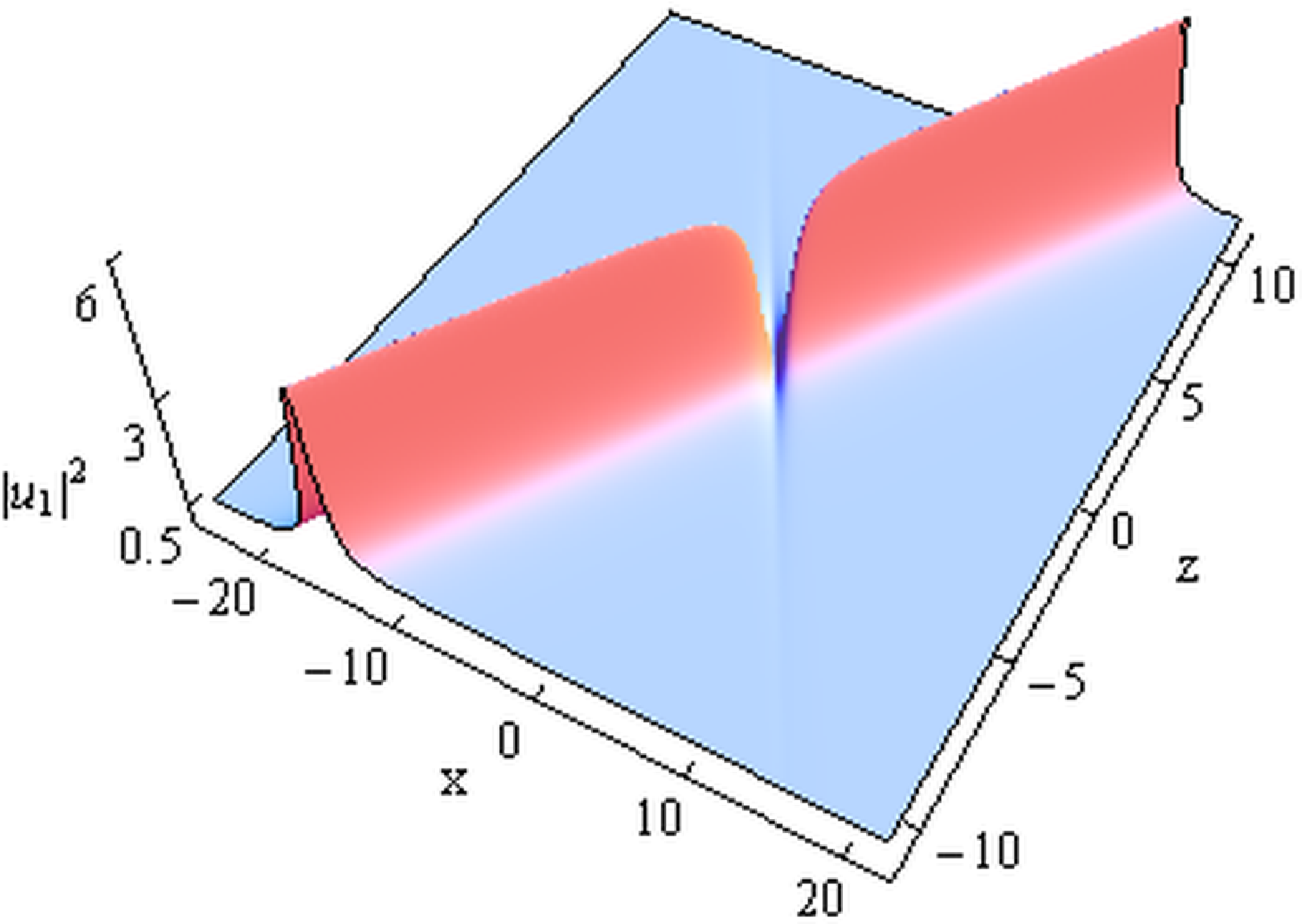}}
\caption{\small Degenerate two-soliton interactions via
the solution in Eq.~\eref{soliton}: (a) Segmented RAD soliton  with
$\rho=\frac{\sqrt{2}}{2}$, $\phi=0$, $\sigma=-1$, $s_1=1$. (b)
Segmented RAD soliton with $\rho=\frac{\sqrt{2}}{2}$, $\phi=0$,
$\sigma=-1$, $s_1=\sqrt{2}\,\mi$. \label{Fig2} }
\end{figure}
Second, we derive the asymptotic expression of
the solution in Eq.~\eref{soliton} along the line $ x - 2\,\sigma \rho z \sim
0$ as $|z| \ra \infty$ as follows:
\begin{subequations}
\begin{align}
& u_1 \ra u^{\rm{I\!I}}_1  := \rho\,\me^{2\,\mi \rho^2 z +\mi
\phi}\bigg[ 1 -\frac{4 \rho (x - 2\,\sigma \rho z) -4\,K^*+4\,\rho
s^*_1 }{2 \rho (x - 2\,\sigma \rho z) -K^*} \bigg], \label{asymp2a}
\\
& |u_1|^2\ra |u^{\rm{I\!I}}_1|^2= \rho^2 \left[1 + \frac{8
\left(\sigma \rho \Im(s_1)+1\right)}{|2 \rho (x - 2\,\sigma
\rho z) -K |^2} \right]. \label{asymp2b}
\end{align}
\end{subequations}
Here, the intensity $|u^{\rm{I\!I}}_1|^2$ can also display the RD
and RAD soliton profiles which are associated with $ 1+\sigma \rho
\Im(s_1) <0$ and $ 1+\sigma \rho \Im(s_1) >0$,
respectively. In this case, both the RD and RAD solitons are
localized along the line $x - 2\,\sigma \rho z  + \Re(s_1)=0 $,
and thus their velocities are given by $v_2 = 2\,\sigma \rho $.  The
height of $|u^{\rm{I\!I}}_1|^2$ for the RAD soliton or the depth of
$|u^{\rm{I\!I}}_1|^2$ for the RD soliton is equal to $A_2 = \frac{8
\rho^2 \left|1+\sigma \rho \Im(s_1)\right|}{\left[\sigma +2
\rho \Im(s_1)\right]^2}$. Similarly, for $ 1+\sigma \rho
\Im(s_1) <0 $, we can use $\omega_2 = \arccos\big[\frac{8
 \left(\rho |\Im(s_1)|- 1 \right)}{\left(2
\rho |\Im(s_1) | -1 \right)^2} \big] $ to characterize the
darkness of the RD soliton. When $ \omega_2 =\frac{\pi}{2}  $ (i.e.,
$ |\Im(s_1)| = \frac{2 \sqrt{2} - 1}{2\,\rho} $),
$u^{\rm{I\!I}}_1$ is a rational black soliton; while for $ 0<
\omega_2 < \frac{\pi}{2} $ (i.e., $ |\Im(s_1)| \neq \frac{2
\sqrt{2} - 1}{2\,\rho}, \frac{1}{\rho}$), it represents a rational
gray soliton.

\begin{table}[h]
\caption{Asymptotic patterns of the solution in Eq.~\eref{soliton} under
different parametric conditions. \label{Table}} \vspace{0mm}
\begin{center}
\begin{tabular}{|c|c|c|c|c|}
\hline  \multicolumn{1}{|c|}{Parametric conditions} & {Asymptotic
soliton $u^{\rm{I}}_1$} & {Asymptotic soliton $u^{\rm{I\!I}}_1$}
\\  \hline  $ \sigma \Im(s_1)< 0$, $ 1+\sigma \rho
\Im(s_1) >0$   &  RAD soliton  &   RAD soliton
\\ \hline
 $ \sigma \Im(s_1) < 0$, $ 1+\sigma \rho
\Im(s_1) < 0$   &  RAD soliton  &   RD soliton
\\ \hline
 $ \sigma \Im(s_1) > 0$, $ 1+\sigma \rho
\Im(s_1) >0$   &  RD soliton  &   RAD soliton
\\ \hline
$  \Im(s_1) = 0$    & Vanish  &   RAD soliton
\\ \hline
$ 1+\sigma \rho \Im(s_1) =0$   &  RAD soliton  &  Vanish
\\ \hline
\end{tabular}
\end{center}
\end{table}

The above asymptotic analysis implies that the solution in Eq.~\eref{soliton}
can describe the elastic interactions of rational solitons in the
sense that two interacting solitons retain their individual shapes,
intensities, and velocities as $z\ra \pm \infty$. However, different from
the standard elastic interaction in the NLS model, each interacting
soliton experiences no phase shift upon an interaction. In general,
the solution in Eq.~\eref{soliton} exhibits three different types of elastic
interactions between two rational solitons on a cw background, as
shown in Fig.~\ref{Fig1}. The associated parametric conditions are
given in the first three rows of Table~\ref{Table}. In particular,
with  $\Im(s_1) = 0$, the asymptotic soliton $u^{\rm{I}}_1$
vanishes as $z\ra\pm \infty$, while $u^{\rm{I\!I}}_1$ displays a
RAD soliton profile [see Fig.~\ref{Fig2a}]. Similarly, for the degenerate
case $1+\sigma \rho \Im(s_1) =0$,  the only surviving asymptotic soliton is
$u^{\rm{I}}_1$  and it takes the shape of the RAD type [see
Fig.~\ref{Fig2b}]. In either of the two degenerate cases, one asymptotic soliton disappears in the far-field region, but it still affects the other one in the near-field region, that is, the surviving soliton is segmented into two pieces at some finite value of $z$. Therefore, such two degenerate cases of the solution in Eq.~\eref{soliton} cannot be simply regarded  as the conventional single soliton~\cite{LiXu}.

\noindent{\em 3.2 Second-order rational soliton solution}

With the truncation of Eqs.~\eref{EP1} and~\eref{EP2} at $j \ge 2$,
we can further obtain a series of higher-order rational soliton solutions.
For $j=2$,  the second-order rational  solution can be obtained as
\begin{align}
\hspace{-10mm}  u_2 = \rho\,\me^{2\,\mi \rho^2 z +\mi\phi} + 2\,\frac{\begin{vmatrix}
f_1^{(0,1)} & f_1^{(1,1)} & f_1^{(2,1)} &  g_1^{(0,1)}  \\
f_1^{(0,3)} & f_1^{(1,3)} & f_1^{(2,3)} &  g_1^{(0,3)} \\
- \bar{g}_1^{(0,1)} &  - \bar{g}_1^{(1,1)} &  - \bar{g}_1^{(2,1)} & \bar{f}_1^{(0,1)}   \\
- \bar{g}_1^{(0,3)} &  - \bar{g}_1^{(1,3)}  &  - \bar{g}_1^{(2,3)} & \bar{f}_1^{(0,3)}
\end{vmatrix}}{
\begin{vmatrix}
f_1^{(0,1)} & f_1^{(1,1)} &  g_1^{(0,1)} &  -g_1^{(1,1)} \\
f_1^{(0,3)} & f_1^{(1,3)} &  g_1^{(0,3)} &  -g_1^{(1,3)} \\
- \bar{g}_1^{(0,1)} &  - \bar{g}_1^{(1,1)} & \bar{f}_1^{(0,1)} & -\bar{f}_1^{(1,1)}  \\
- \bar{g}_1^{(0,3)} &  - \bar{g}_1^{(1,3)} & \bar{f}_1^{(0,3)} & -\bar{f}_1^{(1,3)}
\end{vmatrix}}, \label{2solution}
\end{align}
with
\begin{subequations}
\begin{align}
& f_1^{(0,1)}=2\sqrt{2} \,\mi   \rho\,   \me^{\mi  \rho^2 z +\frac{\mi  \phi }{2}}\zeta, \quad f_1^{(1,1)}=\mi \sigma \rho f_1^{(0,1)}, \quad  f_1^{(2,1)}= -\rho^2 f_1^{(0,1)}, \\
& g_1^{(0,1)}=2 \sqrt{2}  \me^{-\,\mi  \rho^2 z -\frac{\mi  \phi }{2}} \left(\sigma \rho \zeta  + \mi \right),
\quad g_1^{(1,1)}=\mi \sigma \rho g_1^{(0,1)}, \quad  g_1^{(2,1)}= -\rho^2g_1^{(0,1)},
\\
& f_1^{(0,3)}=-\frac{\mi  \rho  \me^{\mi  \rho^2 z +\frac{\mi  \phi }{2}} \left(4 \rho^2 \zeta^3 -3 \zeta -12\,\varpi \right)}{3 \sqrt{2}},  \\
& f_1^{(1,3)}=\frac{\rho^2 \sigma   \me^{\mi  \rho^2 z +\frac{\mi  \phi }{2}} \left(4 \rho^2 \zeta^3 -15 \zeta -12\,\varpi \right)}{3 \sqrt{2}},   \\
& f_1^{(2,3)}=\frac{\mi  \rho^3  \me^{\mi  \rho^2 z +\frac{\mi  \phi }{2}} \left(4 \rho^2 \zeta^3 -27 \zeta -12\,\varpi \right)}{3 \sqrt{2}},   \\
&  g_1^{(0,3)}=-\frac{ \me^{-\,\mi  \rho^2 z -\frac{\mi  \phi }{2}} \left(4 \sigma \rho^3 \zeta^3  +12 \,\mi  \rho^2 \zeta^2 -15 \sigma \rho \zeta -12\,\sigma\rho\,\varpi -3 \,\mi \right)}{3 \sqrt{2}},   \\
& g_1^{(1,3)}=\frac{\sigma \rho \me^{-\,\mi  \rho^2 z -\frac{\mi  \phi }{2}} \left(-4 \,\mi \sigma \rho^3 \zeta^3 +12 \rho^2 \zeta^2 +27 \,\mi  \sigma \rho \zeta + 12\,\mi\,\sigma\rho\,\varpi   -15\right)}{3 \sqrt{2}},   \\
& g_1^{(2,3)}=\frac{\rho^2  \me^{-\,\mi  \rho^2 z -\frac{\mi  \phi }{2}} \left(4\,\sigma \rho^3 \zeta^3  +12 \,\mi  \rho^2 \zeta^2  -39 \sigma  \rho \zeta - 12\,\sigma\rho\,\varpi    -27 \,\mi \right)}{3 \sqrt{2}},
\end{align}
\end{subequations}
where $\bar{f}_1^{(m-1,j)}=f_1^{*(m-1,j)}(-x,z)$, $\bar{g}_1^{(m-1,j)}=g_1^{*(m-1,j)}(-x,z)$ ($1\leq m,j \leq 3 $),
$\zeta=x+2 \,\sigma \rho z +s_1$, $\varpi=2\,\sigma\rho z + s_2$,  $s_1$ and $s_2$ are two complex parameters.

\begin{figure}[H]
 \centering
\subfigure[]{\label{P1a}
\includegraphics[width=2.3in]{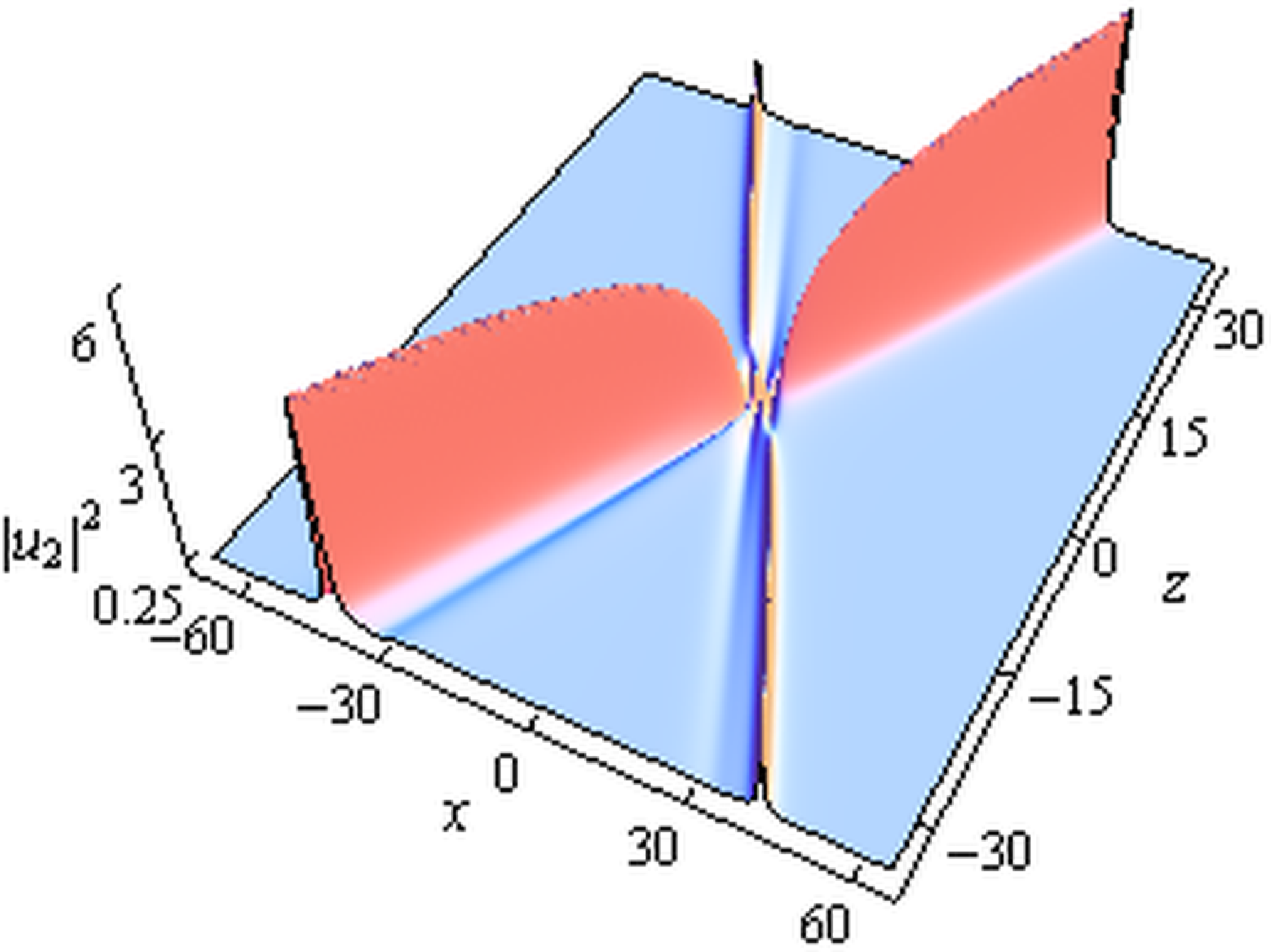}}\hfill
\subfigure[]{ \label{P1b}
\includegraphics[width=3.9in]{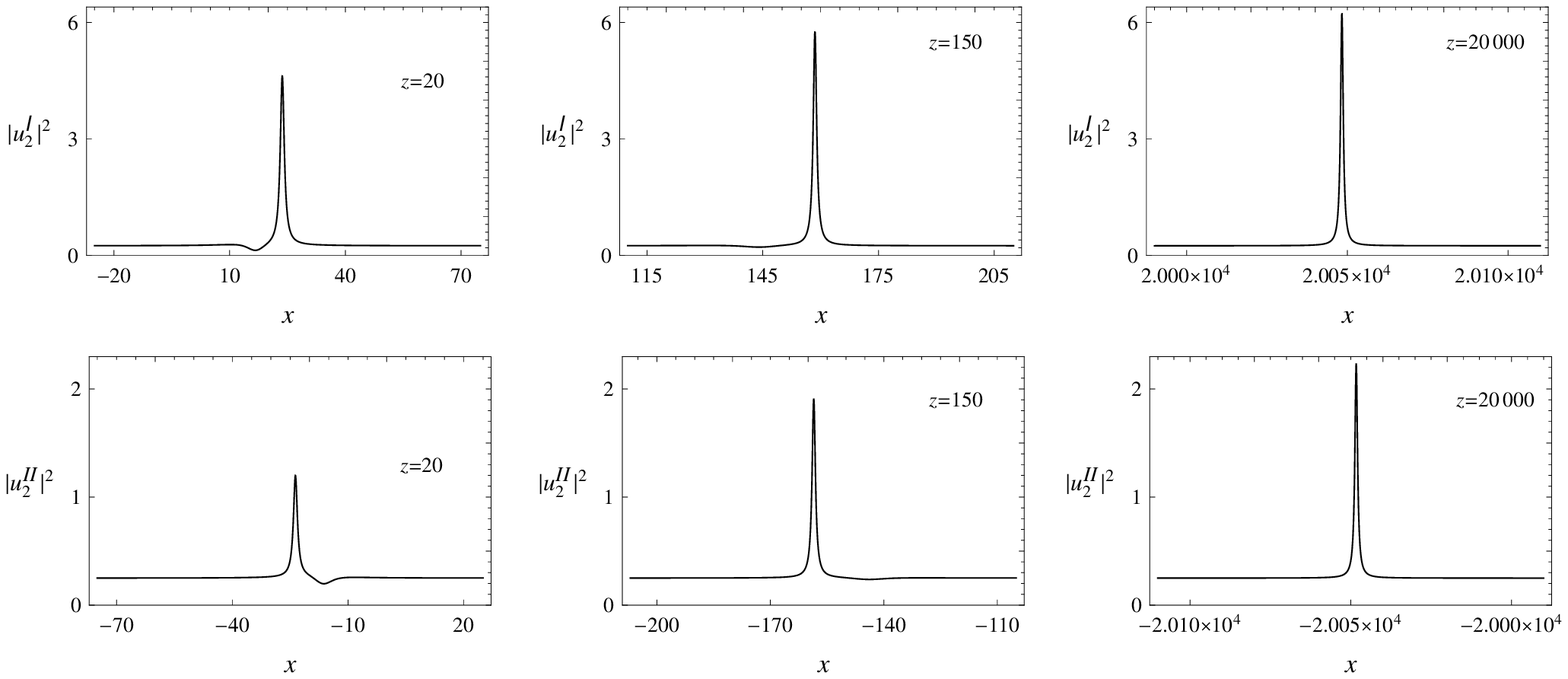}}
\caption{\small First type of soliton interaction via the second-order rational solution in Eq.~\eref{2solution} with $\rho=0.5$, $\phi=0$, $\sigma=-1$, $s_1=1+1.5\,\mi$,
$s_2=1+\mi$.  (a) Rational soliton interaction in the near-field region. (b) Transverse plots of two asymptotic solitons at different values of $z$.  \label{P1} }
\end{figure}

\begin{figure}[H]
 \centering
\subfigure[]{\label{P2a}
\includegraphics[width=2.4in]{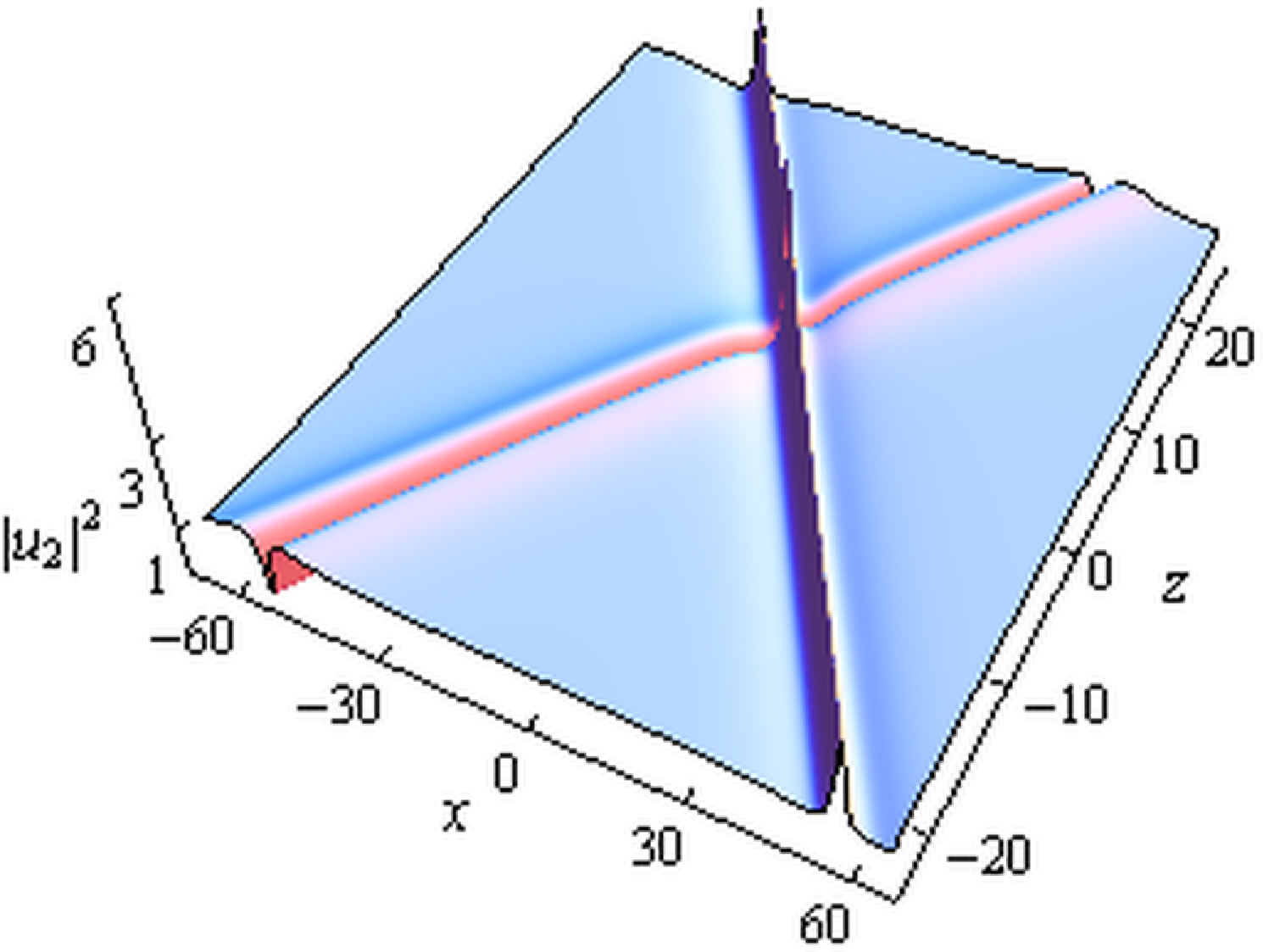}}\hfill
\subfigure[]{ \label{P2b}
\includegraphics[width=3.8in]{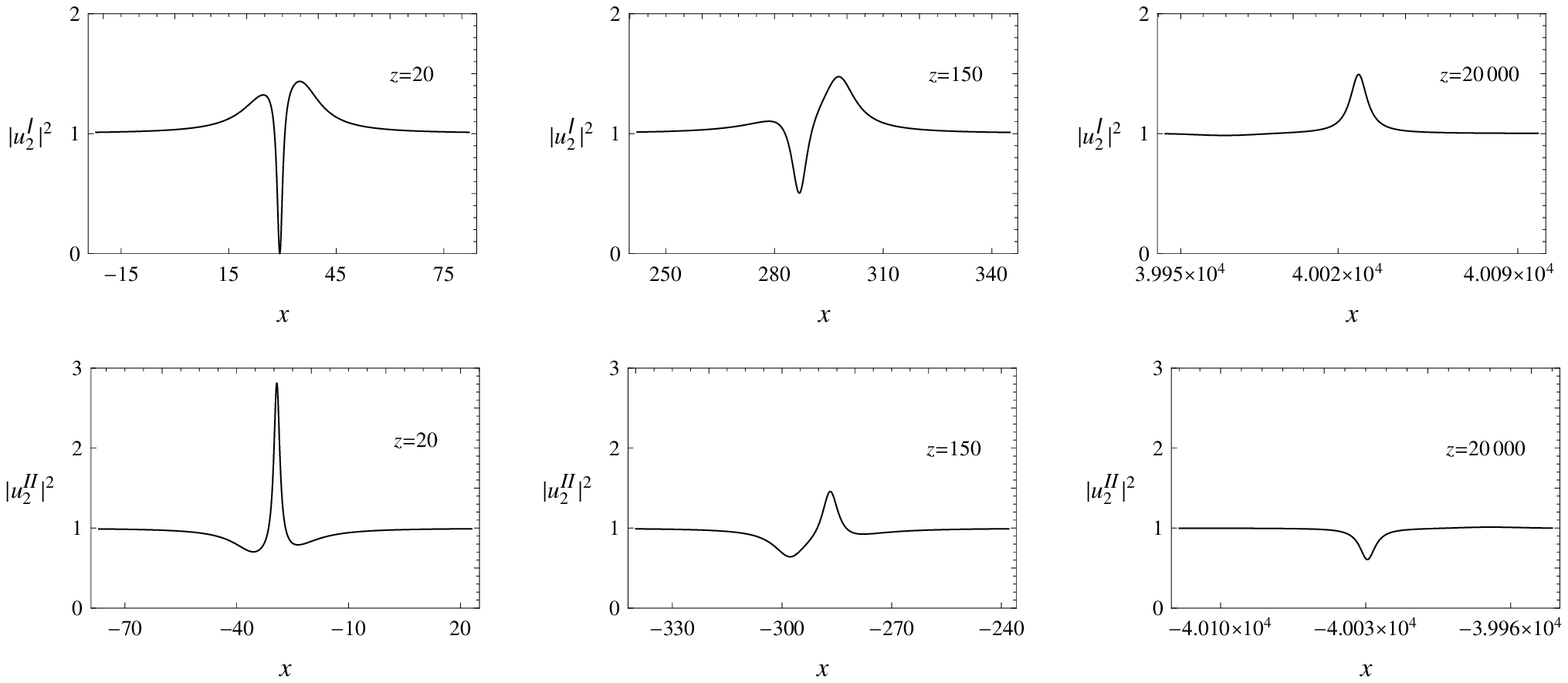}}
\caption{\small Second type of soliton interaction via the second-order rational solution in Eq.~\eref{2solution} with $\rho=1$, $\phi=0$, $\sigma=-1$, $s_1=10 + 5\,\mi$,
$s_2=-4+ 100\,\mi$.  (a) Rational soliton interaction in the near-field region. (b) Transverse plots of two asymptotic solitons at different values of $z$. \label{P2}}
\end{figure}

\begin{figure}[H]
 \centering
\subfigure[]{\label{P3a}
\includegraphics[width=2.4in]{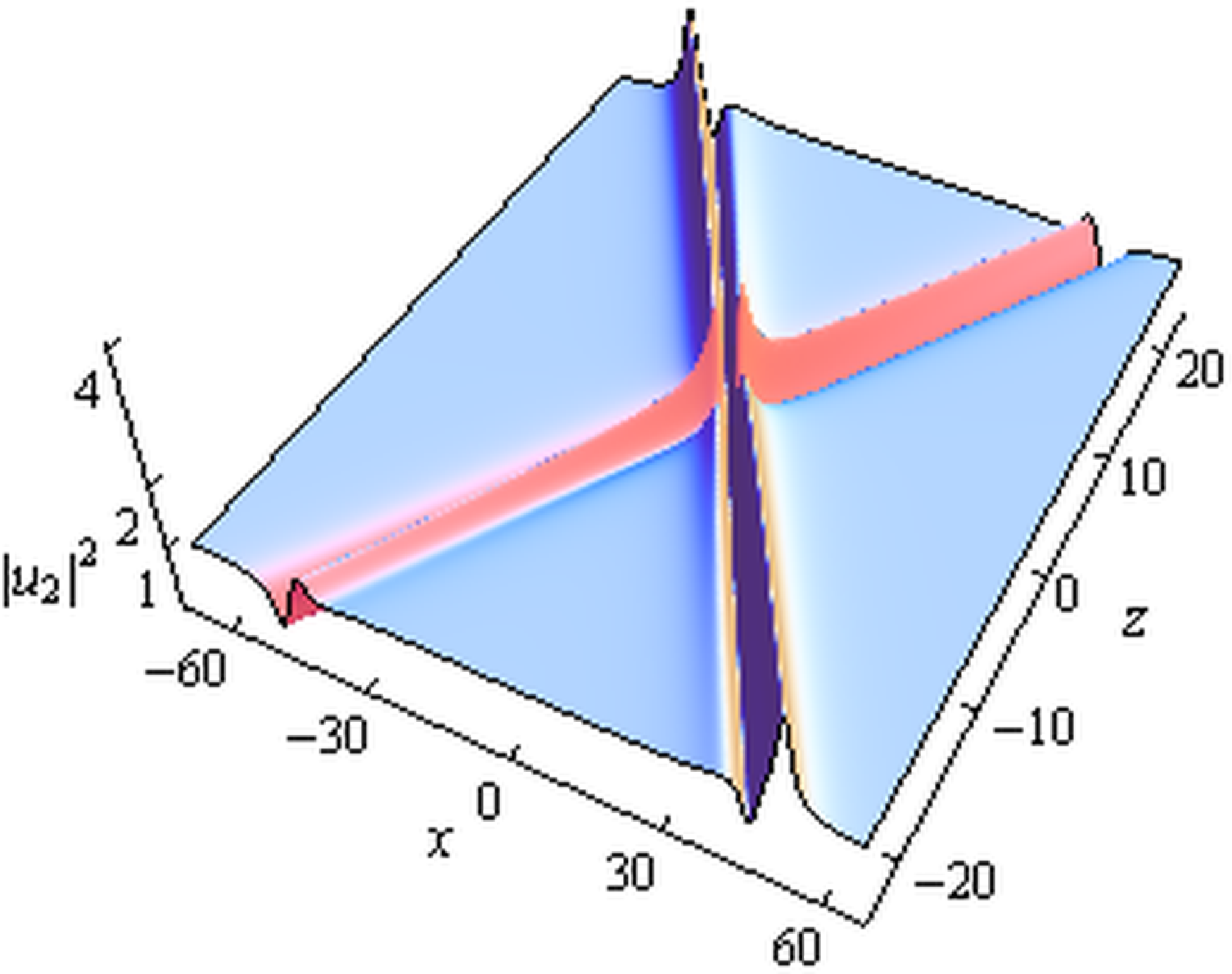}}\hfill
\subfigure[]{ \label{P3b}
\includegraphics[width=3.8in]{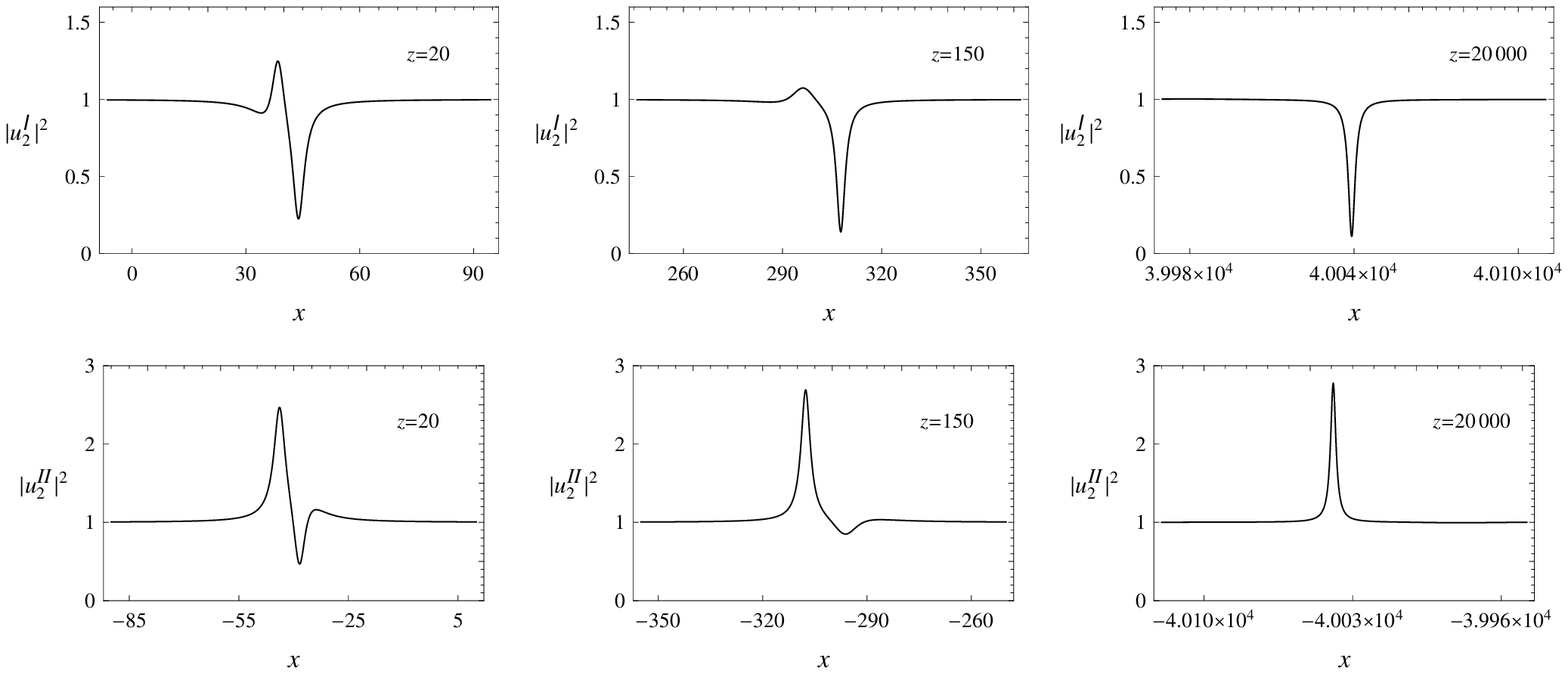}}
\caption{\small Third type of soliton interaction via the second-order rational solution in Eq.~\eref{2solution} with $\rho=1$, $\phi=0$, $\sigma=-1$, $s_1=-\mi$,
$s_2=1-12\,\mi$.  (a) Rational soliton interaction in the near-field region. (b) Transverse plots of two asymptotic solitons at different values of $z$. \label{P3} }
\end{figure}

\begin{figure}[H]
 \centering
\subfigure[]{\label{P4a}
\includegraphics[width=2.4in]{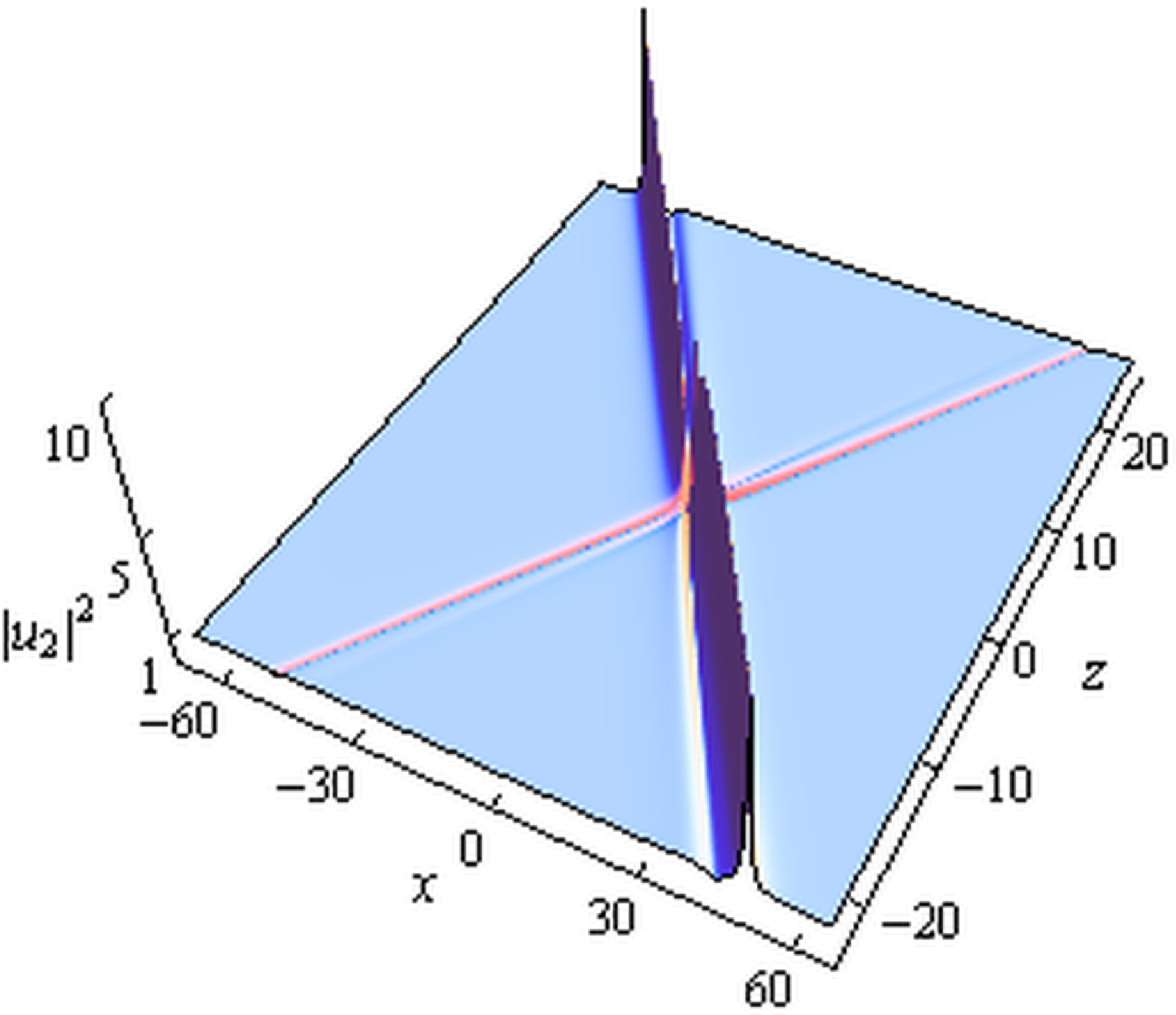}}\hfill
\subfigure[]{ \label{P4b}
\includegraphics[width=3.8in]{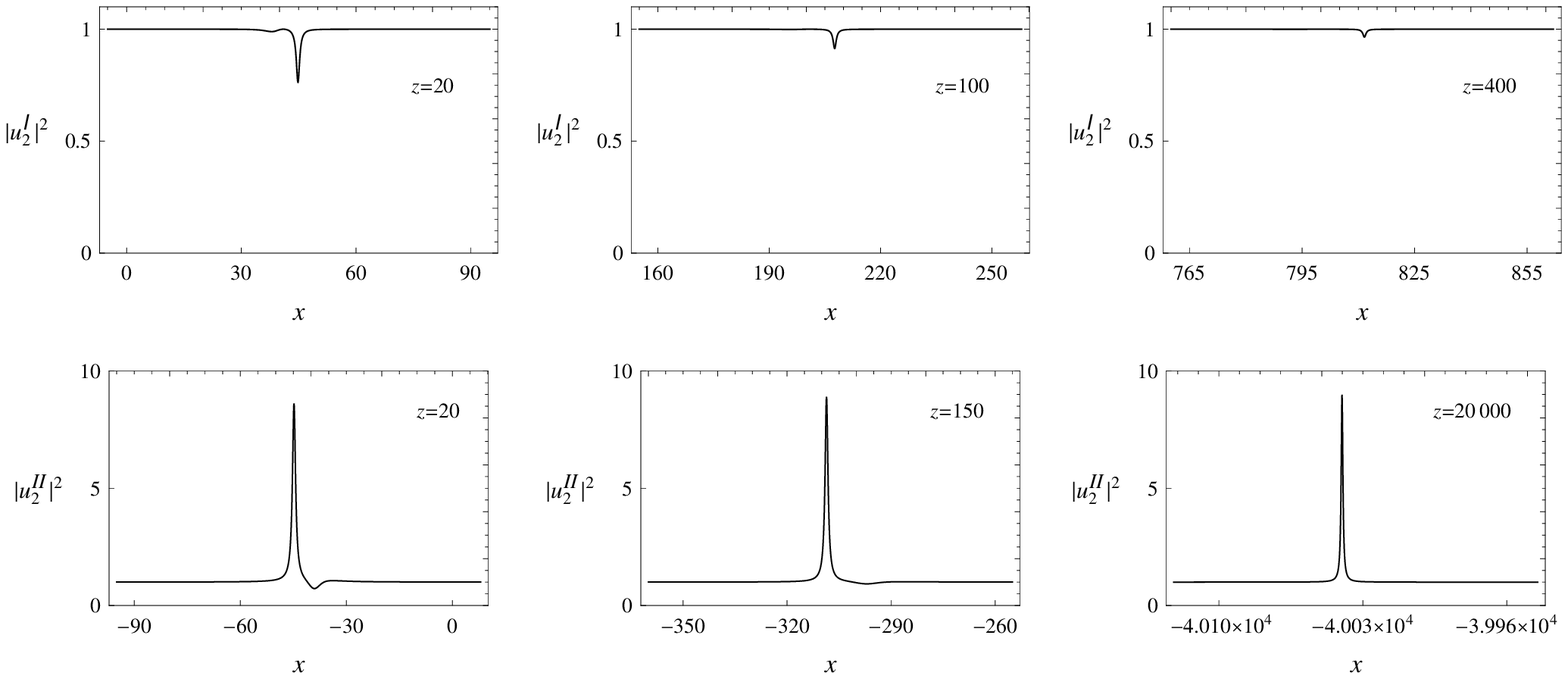}}
\caption{\small First degenerate case of soliton interaction via the second-order rational solution in Eq.~\eref{2solution} with $\rho=1$, $\phi=0$, $\sigma=-1$, $s_1=-1$,
$s_2=1-\mi$.  (a) Rational soliton interaction in the near-field region. (b) Transverse plots of two asymptotic solitons at different values of $z$.  \label{P4}}
\end{figure}

\begin{figure}[H]
 \centering
\subfigure[]{\label{P5a}
\includegraphics[width=2.4in]{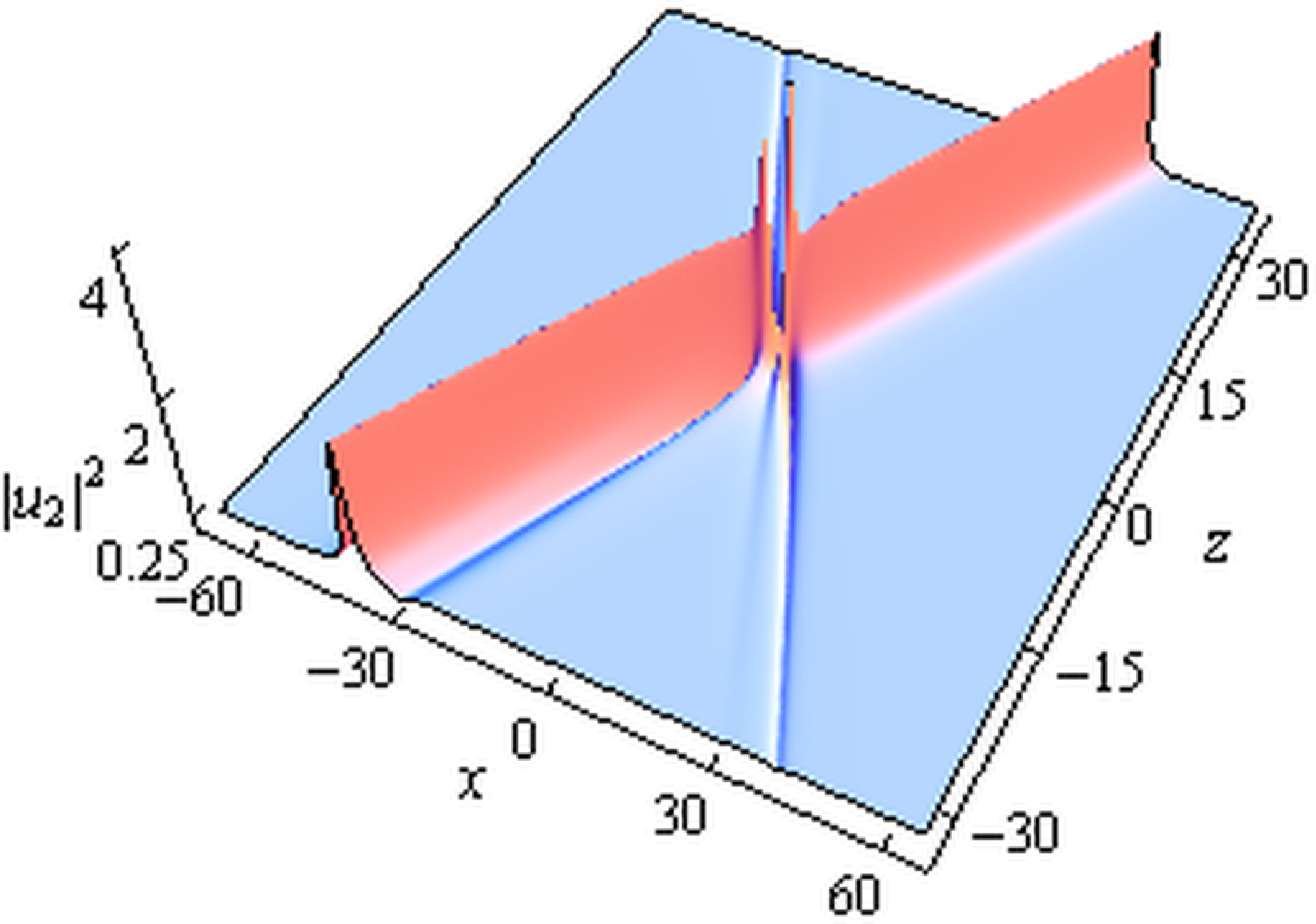}}\hfill
\subfigure[]{ \label{P5b}
\includegraphics[width=3.8in]{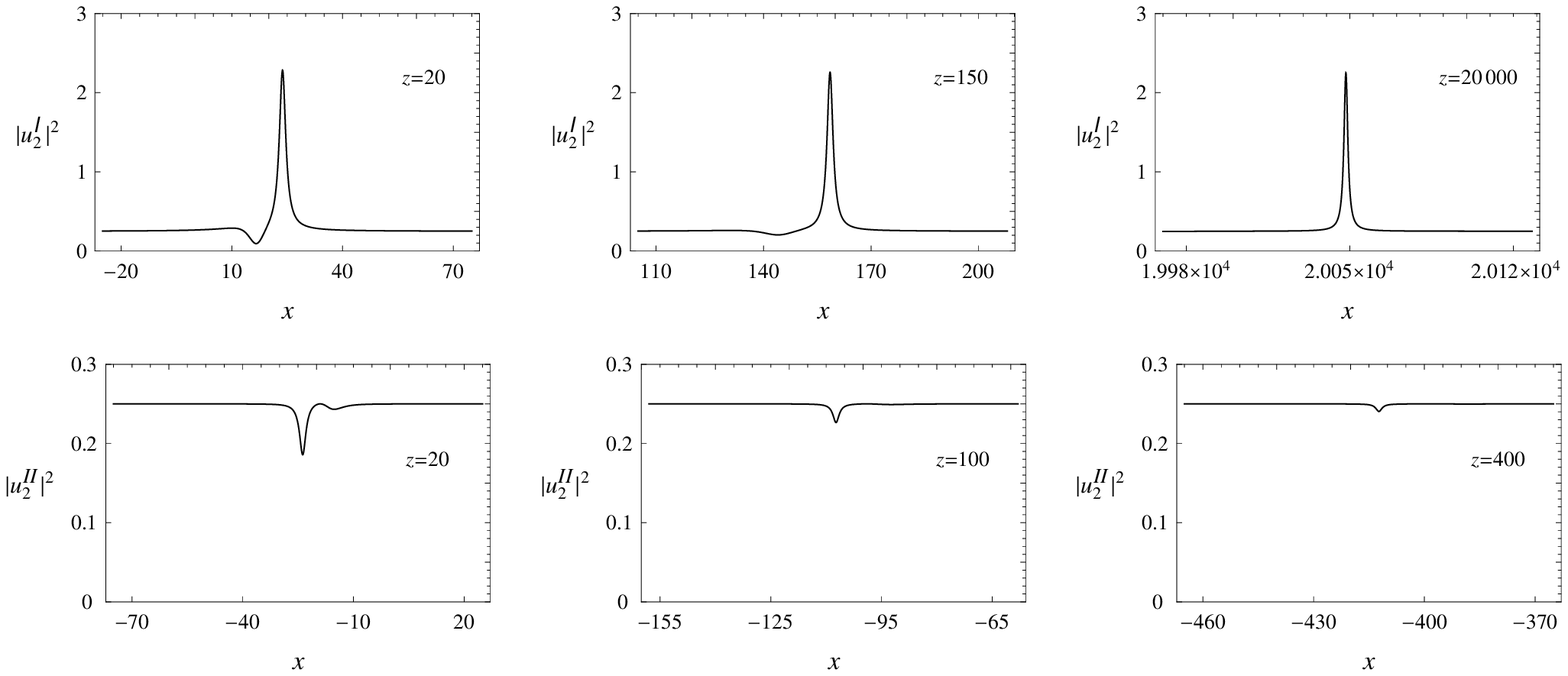}}
\caption{\small Second degenerate  case of soliton interaction via the second-order rational solution in Eq.~\eref{2solution} with $\rho=0.5$, $\phi=0$, $\sigma=-1$, $s_1=1+2\,\mi$,
$s_2=0.2\,\mi$.  (a) Rational soliton interaction in the near-field region. (b) Transverse plots of two asymptotic solitons at different values of $z$.  \label{P5} }
\end{figure}

For the second-order rational solution in Eq.~\eref{2solution}, one can find that it exhibits the finite-amplitude localized wave structures if and only if $s_1$ and $s_2$ satisfy the nonsingular condition
\begin{align}
\left[2 \rho \Im(s_1)+\sigma \right] \left[2 \rho^2
\Im(s_1)^3+3 \rho \sigma  \Im(s_1)^2-3
\Im(s_2)\right]<0.
\end{align}
However, in contrast to the exponential soliton solutions of Eq.~\eref{NNLS}~\cite{LiXu},
the second-order rational solution does not describe the interactions among larger numbers of RD and RAD solitons such as
$u^{\rm{I}}_1$ or $u^{\rm{I\!I}}_1$.
Our asymptotic analysis reveals that there are only two
asymptotic expressions when $ |x - 2\,\sigma \rho z| \ra
\infty $ and $ |x + 2\,\sigma \rho z| \ra \infty$, which
are respectively given as
\begin{align}
&  u_2 \ra u^{\rm{I}}_2 := \rho\me^{2 \mi  \rho^2 z+\mi \phi }
\Big[1 -\frac{6\, \mi\,\sigma \rho (\xi+s_{1})^2  }{2 \rho^2
(\xi+s_{1})^3+3 \,\mi\,\sigma \rho (\xi+s_{1})^2  +3 (2\,\sigma \rho z
+ s_2)} \Big], \label{asymp3} \\
& u_2 \ra u^{\rm{I\!I}}_2 := \me^{2 \mi  \rho^2 z+\mi  \phi}
\Big[\rho - \frac{6\, \mi \sigma  +  12  \rho (\eta-s^*_{1})
-6\,\mi \sigma\rho^2 (\eta-s^*_{1})^2 }{2 \rho^2 (\eta-s^*_{1})^3 +
3\, \mi \,\sigma \rho (\eta-s^*_{1})^2 -  3 (2\,\sigma \rho z +
s^*_2) } \Big], \label{asymp4}
\end{align}
with  $\xi  $ and  $\eta $ defined below Eq.~\eref{soliton}. By a numerical comparison, it can be verified that the solution in Eq.~\eref{2solution} agrees very well with Eqs.~\eref{asymp3} and~\eref{asymp4} at large values of $z$. Through
the qualitative analysis of Eqs.~\eref{2solution}, \eref{asymp3} and~\eref{asymp4} and the results in Figs.~\ref{P1}--\ref{P5}, we find that
the second-order rational solution has the following properties:
\begin{enumerate}
\item[(i)]
In the near-field region $|z|\ll\infty$, the intensity profiles of $u^{\rm{I}}_2 $ and $u^{\rm{I\!I}}_2 $
display various combined-peak-valley soliton structures, which contain one or two peak(s) and valley(s)
and vary with the evolution of $z$.  As  $|z|\ra \infty$, either $u^{\rm{I}}_2 $ or $u^{\rm{I\!I}}_2 $
will \emph{slowly} evolve into a RAD soliton or a RD soliton  if $ \Im(s_1) \neq 0$ and $ 1+\sigma \rho \Im(s_1) \neq 0$ [see Figs.~\ref{P1b}--\ref{P3b}], and $u^{\rm{I}}_2 $ (or $u^{\rm{I\!I}}_2 $) will  vanish if
$ \Im(s_1) = 0$ (or $ 1+\sigma \rho \Im(s_1) = 0$) [see Figs.~\ref{P4b} and~\ref{P5b}].

\item[(ii)]

The combined-peak-valley solitons $u^{\rm{I}}_2 $ and $u^{\rm{I\!I}}_2 $ interact elastically in the near-field regions [as shown in Figs.~\ref{P1a}--\ref{P5a}], that is, the shapes, amplitudes, velocities, and phases of two interacting solitons remain the same  after  their mutual interactions.

\item[(iii)]

The parameter $s_1$ determines whether $u^{\rm{I}}_2 $ and $u^{\rm{I\!I}}_2 $ will eventually vanish or evolve into a RD  or RAD soliton as  $|z|\ra \infty$,  while $s_2$ only affects the peak height and the valley depth for each interacting soliton in the near-field regions.

\item[(iv)]

According to the ultimate states of $u^{\rm{I}}_2 $ and $u^{\rm{I\!I}}_2 $ as  $|z|\ra \infty$, the second-order rational solution can exhibit three different types of elastic interactions (Figs.~\ref{P1}--\ref{P3}) and two  degenerate cases (Figs.~\ref{P4} and~\ref{P5}). Moreover, the parametric conditions of the five cases are the same as those listed in Table~\ref{Table} for the first-order rational solution~\eref{soliton}.

\end{enumerate}

\noindent{\em 3.3 Stability analysis via numeric simulation}

\begin{figure}[H]
 \centering
\subfigure[]{\label{Fig8a}
\includegraphics[width=2.1in]{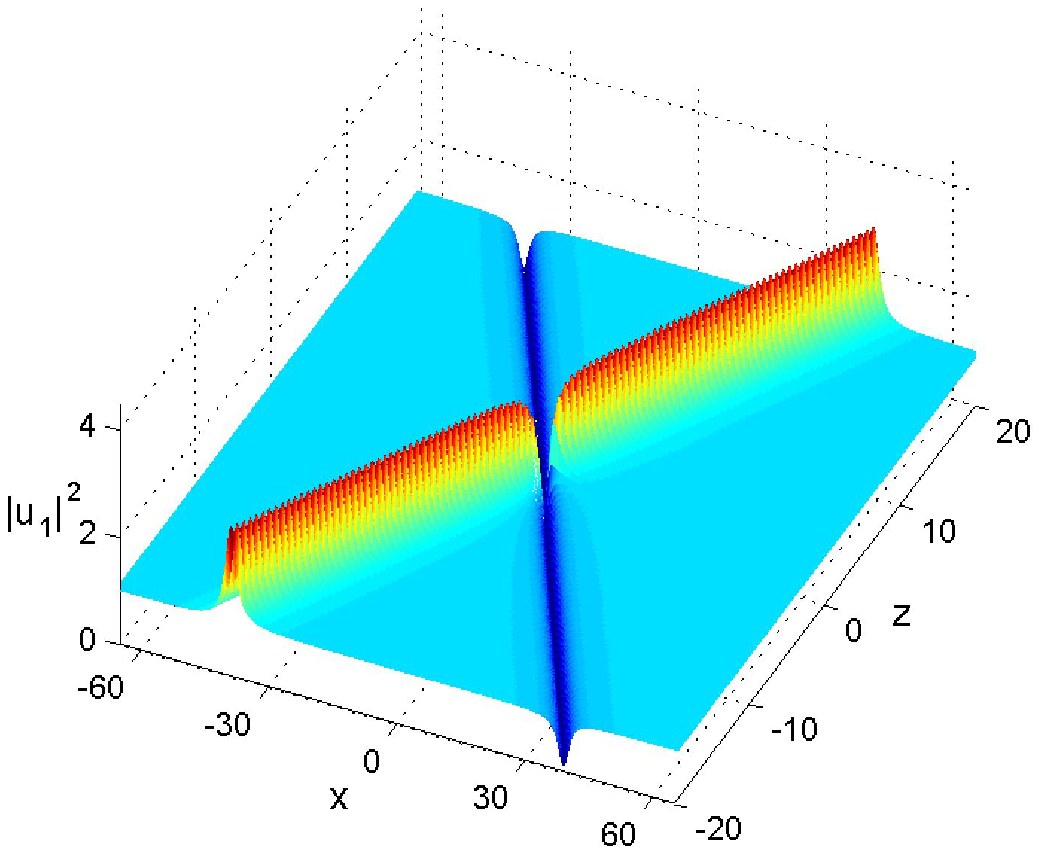}}\hfill
\subfigure[]{ \label{Fig8b}
\includegraphics[width=2.1in]{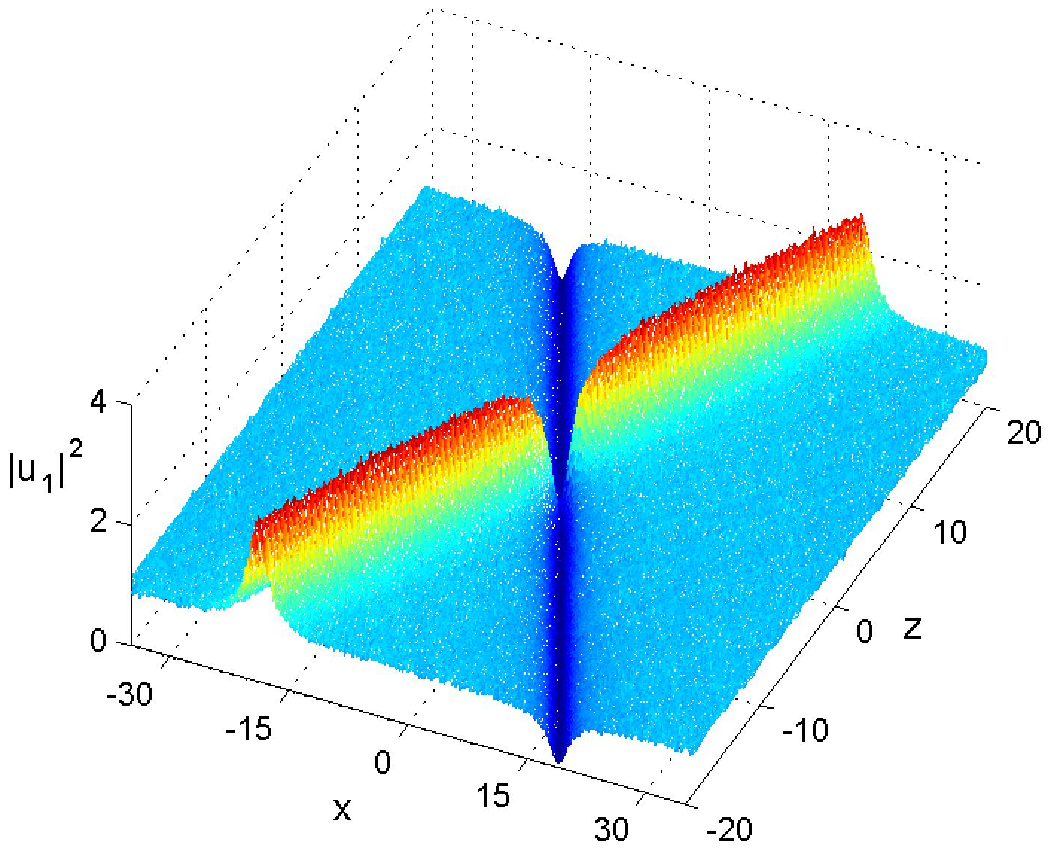}}\hfill
\subfigure[]{ \label{Fig8c}
\includegraphics[width=2.1in]{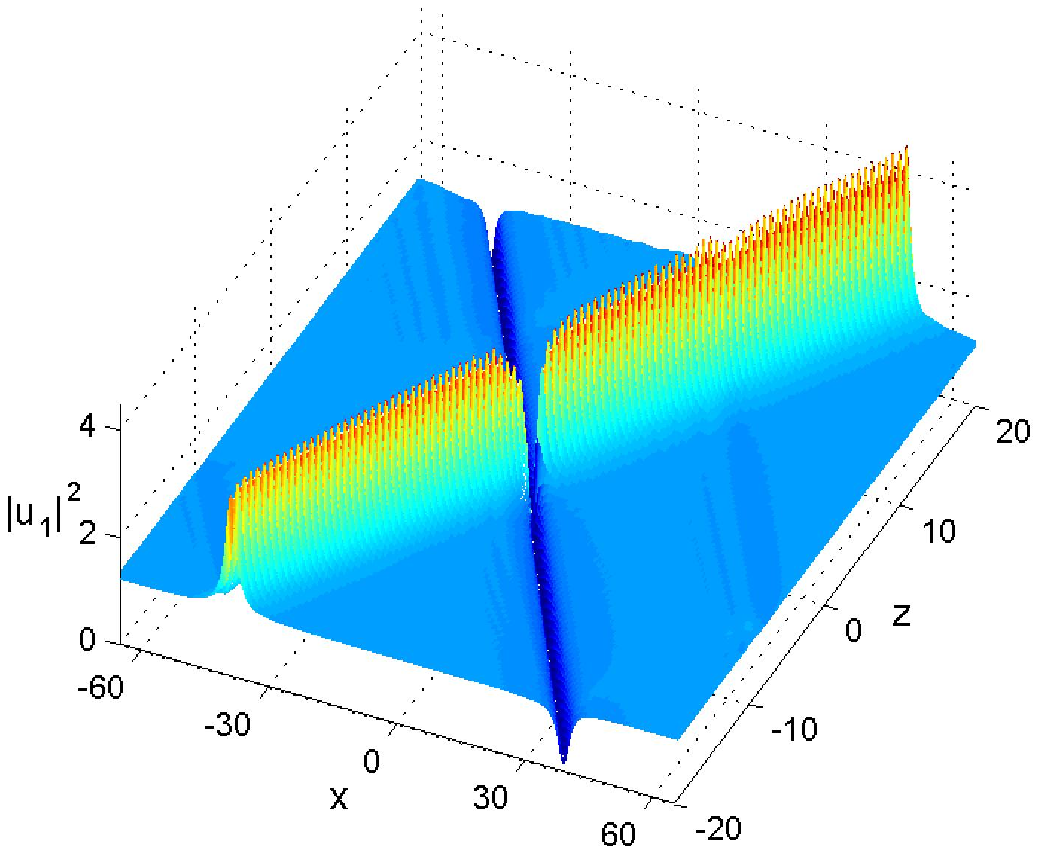}}\\
\subfigure[]{\label{Fig8d}
\includegraphics[width=2in]{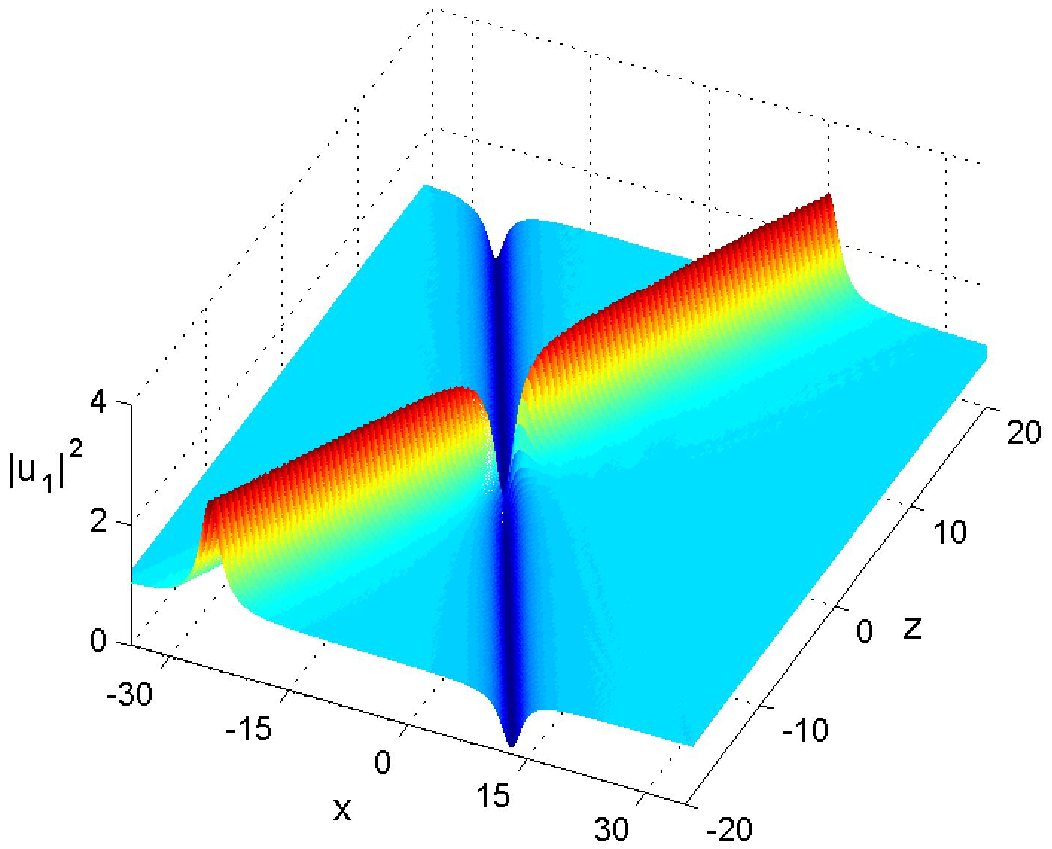}}\hspace{2.0cm}
\subfigure[]{ \label{Fig8e}
\includegraphics[width=2.1in]{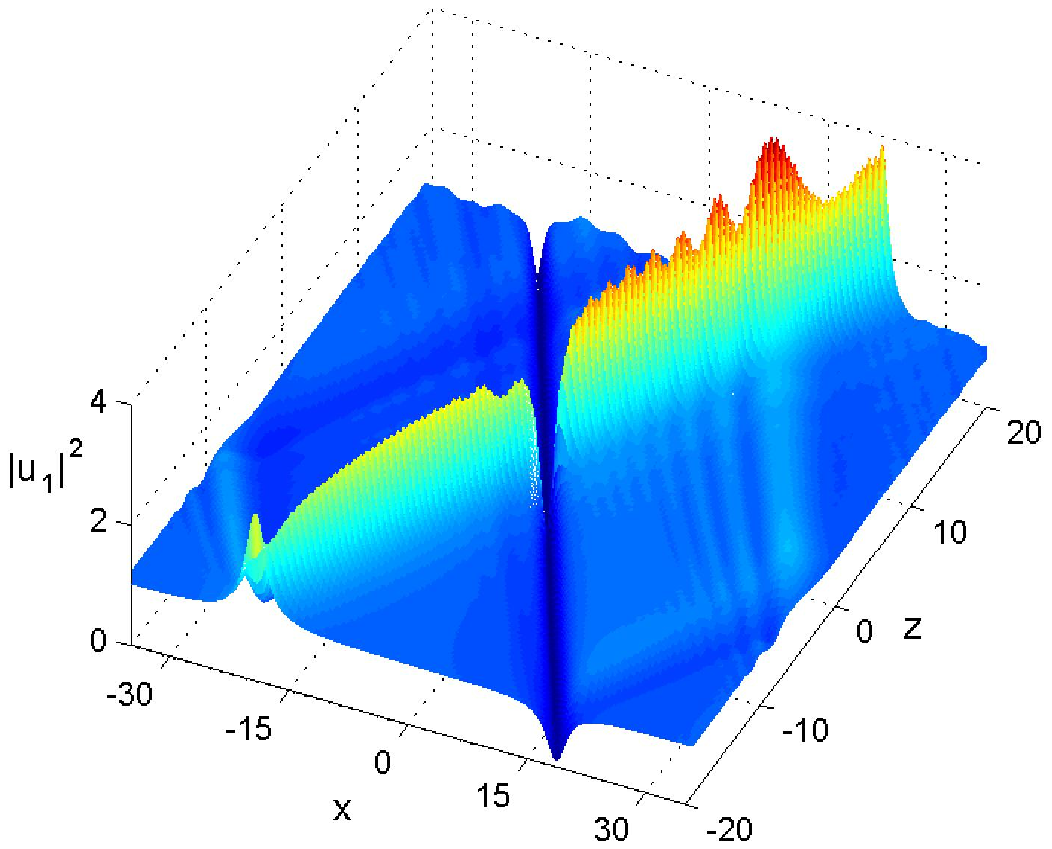}}
\caption{\small Numerical simulations of the first-order rational soliton solution.
(a) The initial value is the exact solution in Eq.~\eref{soliton} at $z=-20$ with $\rho=1$, $\phi=0$, $\sigma=1$,
$s_1=1+\mi$. (b) A white noise with a maximal value of $0.1$ is added to the initial value.  (c) The amplitude of the initial value is amplified by $10\%$. (d) $u(x,-20)$ and $u^*(-x,-20)$ have the same shift $x_0=6$. (e) $u(x,-20)$ and $u^*(-x,-20)$ have the opposite shift $x_0=0.3$.}  \label{Fig8}
\end{figure}
\begin{figure}[H]
 \centering
\subfigure[]{\label{Fig9a}
\includegraphics[width=2.1in]{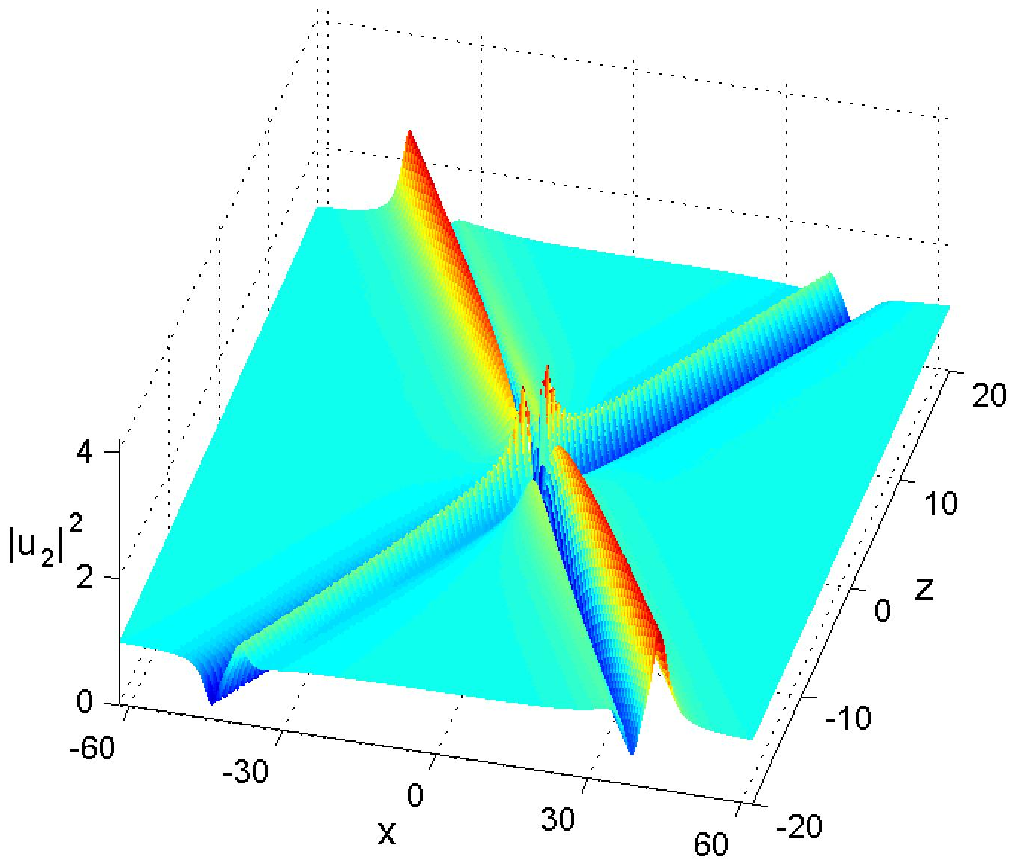}}\hfill
\subfigure[]{ \label{Fig9b}
\includegraphics[width=2.1in]{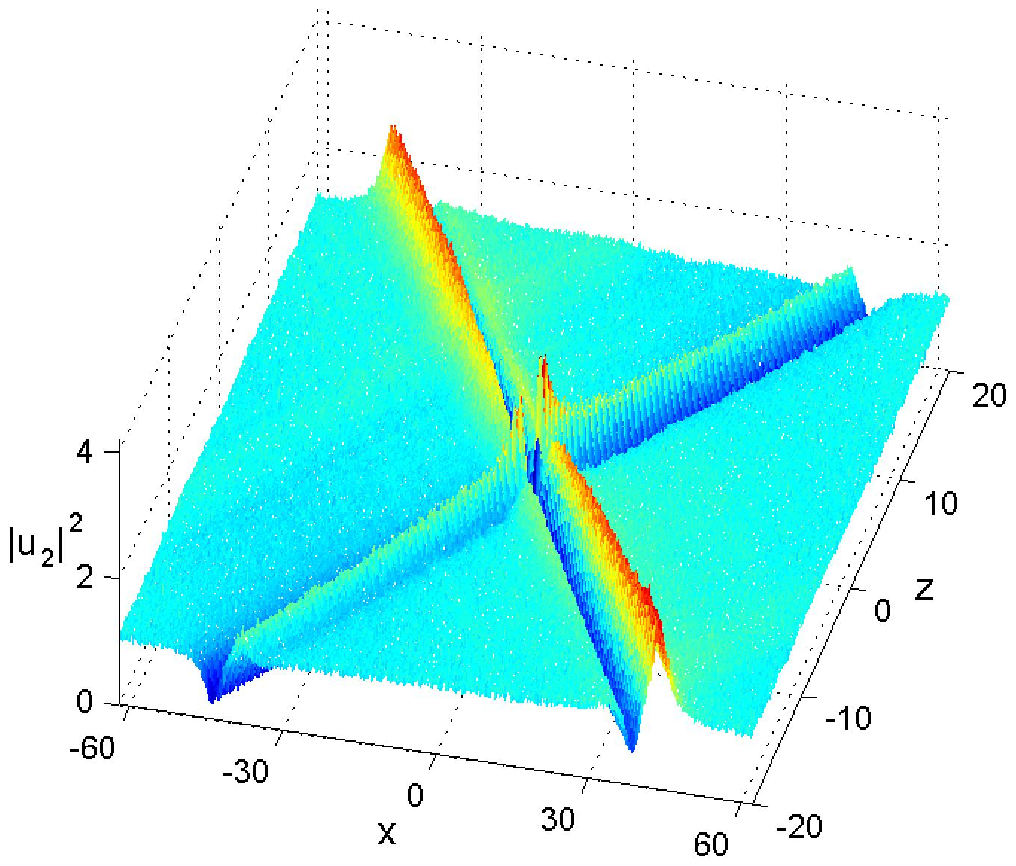}}\hfill
\subfigure[]{ \label{Fig9c}
\includegraphics[width=2.1in]{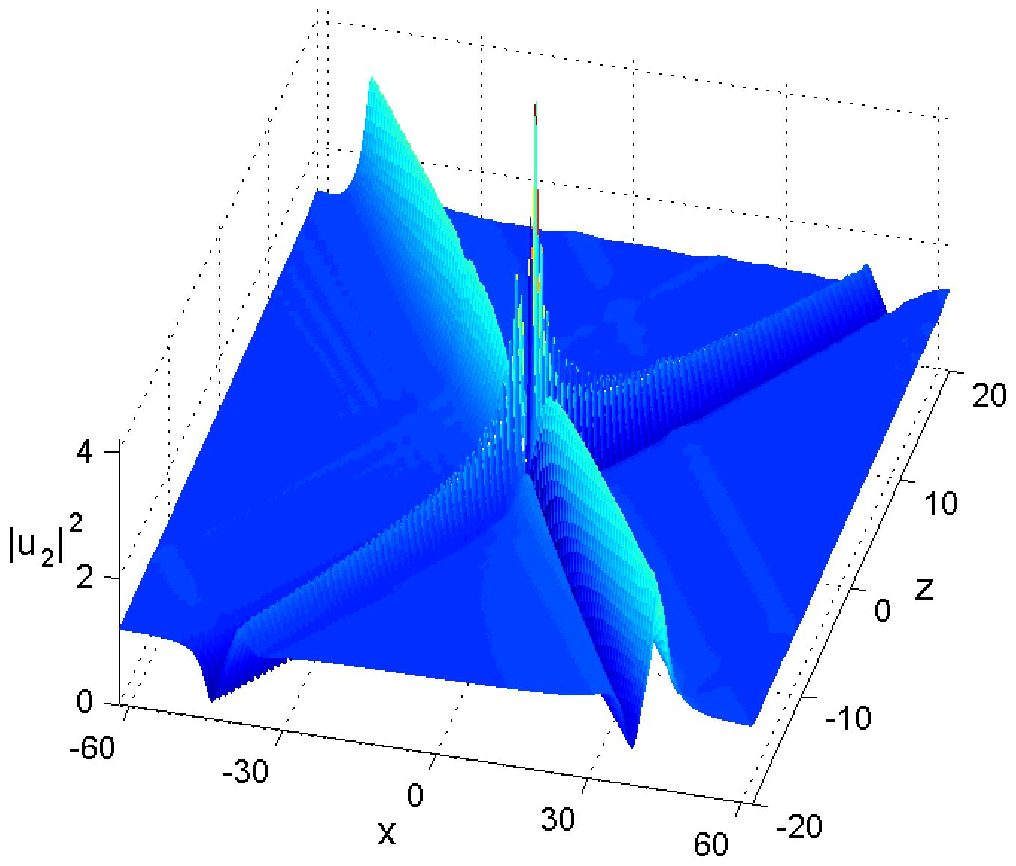}}\\
\subfigure[]{\label{Fig9d}
\includegraphics[width=2in]{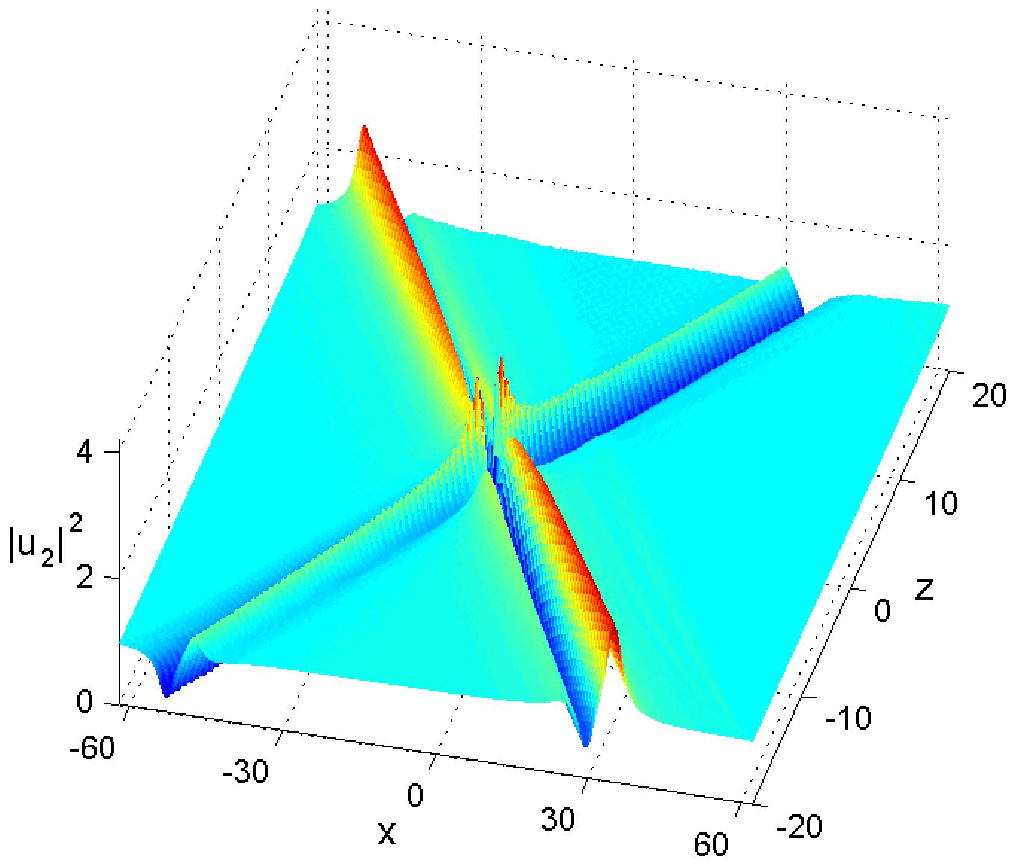}}\hspace{2.5cm}
\subfigure[]{ \label{Fig9e}
\includegraphics[width=2.1in]{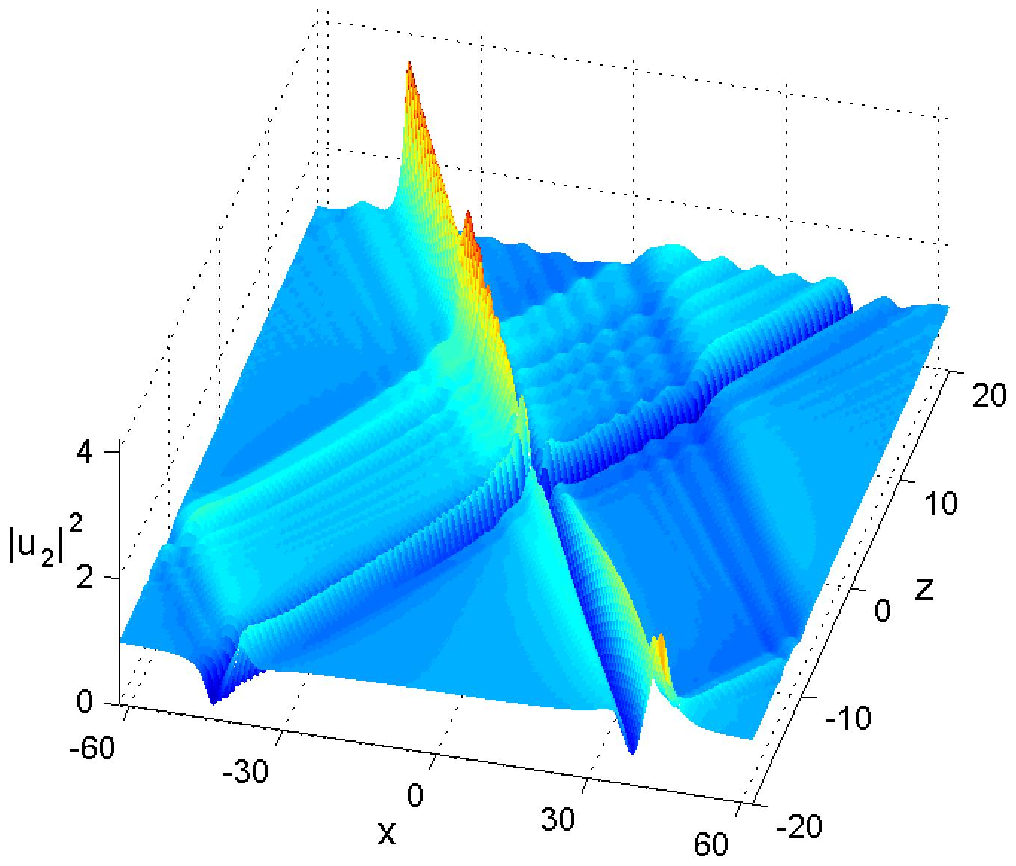}}
\caption{\small Numerical simulations of the second-order rational soliton solution.
(a) The initial value is the exact solution in Eq.~\eref{2solution} at $z=-20$ with $\rho=1$, $\phi=0$, $\sigma=-1$, $s_1=-\mi$, $s_2=1-12\,\mi$. (b) A white noise with a maximal value of $0.1$ is added to the initial value.  (c) The amplitude of the initial value is amplified by $10\%$. (d) $u(x,-20)$ and $u^*(-x,-20)$ have the same shift $x_0=9$. (e) $u(x,-20)$ and $u^*(-x,-20)$ have the opposite shift $x_0=0.2$.}  \label{Fig9}
\end{figure}

On one hand, we use the time-splitting Fourier method to study the stability of the rational soliton solutions with respect to finite initial perturbations. We choose the solutions in Eqs.~\eref{soliton} and~\eref{2solution} at $z=-20$ as the initial values, and examine two types of initial perturbations: (i)  small white noise on  the initial values and (ii) small perturbation on the initial amplitude.  In Figs.~\ref{Fig8a} and~\ref{Fig9a},  the numerical simulations  present the stable evolution of the first- and second-order rational soliton solutions with the same parameters as those in Figs.~\ref{Fig1c} and~\ref{P3}, respectively.  When a white noise with a maximal value of 0.1 is added to the initial values, it can be seen from Figs.~\ref{Fig8b} and~\ref{Fig9b} that the propagations of the first- and second-order rational soliton solutions are affected very little by the white noise. If the amplitudes of the initial values are amplified by $10\%$, Figs.~\ref{Fig8c} and~\ref{Fig9c} show that the magnitudes of two interacting solitons are enhanced but the soliton shapes are  maintained very well.

On the other hand, we note that the exact solution of Eq.~\eref{NNLS} can form a  \PT-symmetric self-induced potential that maintains the stable localized soliton structures. Here, we are concerned with whether the \PT-symmetry breaking of the self-induced potential will lead to the instability of rational solitons. First, we consider that the initial values of $u(x,z)$ and $u^*(-x,z)$ have the same shift in the $x$-coordinate, i.e., $u(x,z)\ra u(x-x_0,z),\,u^*(-x,z)\ra u^*(-x+x_0,z)$. In this case, there is no instability occurring in the evolution of the first- and second-order rational soliton solutions [see Figs.~\ref{Fig8d} and~\ref{Fig9d}], although the symmetric center of the self-induced potential is shifted to $x_0$. The reason lies in the fact that Eq.~\eref{NNLS} remains invariant under the coordinate transformation $x \ra x+x_0$. Second, for the opposite  shift in the $x $-axis  $u(x,z)\ra u(x-x_0,z),\,u^*(-x,z)\ra u^*(-x-x_0,z)$,  the self-induced potential cannot keep the \PT-symmetry with respect to any point of $x$. Hence, an obvious instability will appear  for both the first- and second-order rational soliton solutions, as displayed in Figs.~\ref{Fig8e} and~\ref{Fig9e}. As the value of $x_0$ increases, the soliton solutions  become more unstable and the localized structures are  finally destroyed. This is similar to the situation for the exponential soliton solutions of Eq.~\eref{NNLS}~\cite{LiXu}.

\section{Conclusions} \label{Sec4}

We have studied the rational nonlinear localized wave
phenomena on the cw background for the \PT-symmetric nonlocal NLS
model with the defocusing-type nonlinearity. By the
generalized DT, we have derived the nonsingular rational soliton
solutions starting from a cw solution. For the first-order rational
solution, we have revealed the elastic RAD-RAD, RD-RAD, and RAD-RD
soliton interactions, in which there is no phase shift for interacting
solitons. Meanwhile, we have discussed the degenerate case in which only one RD or RAD soliton survives. We have found that the second-order rational soliton solution does not exhibit
the elastic interactions among larger numbers of fundamental rational solitons such as $u^{\rm{I}}_1$ in Eq.~\eref{asymp1a} or $u^{\rm{I\!I}}_1$ in Eq.~\eref{asymp2a}, but the two interacting solitons display a rich variety of combined-peak-valley structures in the near-field regions, and each of them eventually vanishes or evolves into a RD  or RAD soliton as  $|z|\ra \infty$. Also, we have numerically analyzed the stability of the first- and second-order rational soliton solutions.
The results show that the soliton structures are stable with the addition of a white noise to the initial value and under small perturbations of the amplitude of the initial value, but the stability will be destroyed if  the  self-induced potential loses the
\PT~symmetry  with respect to any point of $x$.

\section*{Acknowledgements}
This work was supported by the National Natural Science Foundation of China (Grant Nos. 61505054, 11426105, and 11326143), by the Fundamental Research Funds of
the Central Universities (Project Nos. 2014QN30, 2014ZZD10, and
2015ZD16),  and by the Science Foundations of China
University of Petroleum, Beijing (Grant Nos. 2462015YQ0604 and YJRC02013-16).
The author T. Xu thanks Prof. J. He for sharing his study on the rational solutions of Eq.~\eref{NNLS} in the 2014 SIAM conference on ``Nonlinear Waves and Coherent Structures''.


\begin{thebibliography}{99}

\bibitem{Shankar}
R.\ Shankar, {\em Principles of Quantum Mechanics} (Plenum Press, New
York, 1994).

\bibitem{bender1}
C.\ M.\ Bender and S.\ Boettcher, {Phys.\ Rev.\ Lett.} {\bf 80},
5243 (1998).

\bibitem{bender2}
C.\ M.\ Bender, D. C. Brody, and H. F. Jones,  {Phys. Rev. Lett.}
{\bf 89}, 270401 (2002).

\bibitem{bender3}
C.\ M.\ Bender, D.\ C.\ Brody, H.\ F.\ Jones, and B.\ K.\ Meister,
{Phys.\ Rev.\ Lett.} {\bf98}, 40403 (2007).

\bibitem{Lie}
B. Bagchi and C. Quesne,
{Phys. Lett. A} {\bf 273}, 285 (2000). 

\bibitem{crystal}
S. Longhi, 
{Phys. Rev. Lett.}  {\bf 103}, 123601 (2009).      

\bibitem{Markum}
H. Markum, R. Pullirsch, and T. Wettig,
{Phys. Rev. Lett.} {\bf 83}, 484 (1999). 

\bibitem{BEC}
H. Cartarius and G. Wunner, {Phys. Rev. A} {\bf 86}, 013612 (2012).

\bibitem{mechanics}
C. M. Bender, B. K. Berntson, D. Parker, and E. Samuel,
{Am. J. Phys.} {\bf 81}, 173 (2013). 

\bibitem{OL2007}
R. El-Ganainy, K. G. Makris, D. N. Christodoulides, and Z. H.
Musslimani, 
Opt. Lett. {\bf 32}, 2632 (2007). 


\bibitem{Makris1} K.\ G.\ Makris, R.\ El-Ganainy,  D.\ N.\
Christodoulides, and Z. H. Musslimani, {Phys.\ Rev.\ Lett.}\ {\bf 100}, 103904 (2008).

\bibitem{Makris2}K. G. Makris, R. El-Ganainy, D. N. Christodoulides, and Z.
H. Musslimani, 
{Phys. Rev. A} {\bf 81}, 063807 (2010).  

\bibitem{Lin}Z. Lin, H. Ramezani, T. Eichelkraut, T. Kottos, H. Cao, and D.
N. Christodoulides, 
{Phys. Rev. Lett.}  {\bf 106}, 213901 (2011).


\bibitem{Guo}A. Guo, G. J. Salamo, D. Duchesne, R. Morandotti, M.
Volatier-Ravat, V. Aimez, G. A. Siviloglou, and D. N.
Christodoulides, 
{Phys. Rev. Lett.} {\bf 103}, 093902 (2009).

\bibitem{Ruter}
C. E. R\"{u}ter, K. G. Makris, R. El-Ganainy, D. N.
Christodoulides, M. Segev, and D. Kip, 
{Nat. Phys.}  {\bf 6}, 192 (2010). 

\bibitem{Castaldi}
G. Castaldi, S. Savoia, V. Galdi, A. Alu, and N. Engheta,
{Phys. Rev. Lett.} {\bf 110}, 173901 (2013).

\bibitem{Regensburger}
A. Regensburger, C. Bersch, M.-A. Miri, G.
Onishchukov, D. N. Christodoulides, and U. Peschel,
Nature {\bf 488}, 167 (2012).  

\bibitem{Musslimani1} Z.\ H.\ Musslimani, K.\ G.\ Makris, R.\ El-Ganainy, and D.\ N.\ Christodoulides,
{Phys.\ Rev.\ Lett.}\ {\bf100}, 030402 (2008).

\bibitem{Gauss}
S. Hu, X. Ma, D. Lu, Z. Yang, Y. Zheng, and W. Hu,
{Phys. Rev. A} {\bf 84}, 043818 (2011).

\bibitem{Harmonic}
D. A. Zezyulin and V. V. Konotop, 
{Phys. Rev. A} {\bf 85}, 043840 (2012).

\bibitem{RosenMorse}
B. Midya and R. Roychoudhury, 
{Phys. Rev. A}  {\bf 87}, 045803 (2013).

\bibitem{Gap}
C. Li, C. Huang, H. Liu, and L. Dong, {Opt. Lett.} {\bf 37}, 4543
(2012); S. Liu, C. Ma, Y. Zhang, and K. Lu, {Opt. Commun.} {\bf
285}, 1934 (2012).

\bibitem{Defect}
H. Wang and J. Wang, {Opt. Express} {\bf 19}, 4030 (2011); Z. Lu and
Z. Zhang, {Opt. Express} {\bf 19}, 11457 (2011).

\bibitem{NonlinearLattice}F. K. Abdullaev, Y. V. Kartashov, V. V. Konotop, and D.
A. Zezyulin, {Phys. Rev. A} {\bf 83}, 041805(R) (2011);  D. A.
Zezyulin, Y. V. Kartashov, and V. V. Konotop,  EPL {\bf 96}, 64003
(2011).

\bibitem{MixLattice}
Y. He, X. Zhu, D. Mihalache, J. Liu, and Z. Chen,
Phys. Rev. A  {\bf 85}, 013831 (2012).


\bibitem{Vector}
Y. V. Kartashov, 
Opt. Lett. {\bf 38}, 2600 (2013).

\bibitem{Breather}
I. V. Barashenkov, S. V. Suchkov, A. A. Sukhorukov, S. V. Dmitriev,
and Y. S. Kivshar, 
Phys. Rev. A  {\bf 86}, 053809 (2012).

\bibitem{Rogue}C. Q. Dai and W. H. Huang,
{Appl. Math. Lett.} {\bf 32}, 35 (2014).

\bibitem{Lumer}
Y. Lumer, Y. Plotnik, M. C. Rechtsman, and M. Segev,
{Phys. Rev. Lett.} {\bf 111}, 263901 (2013).

\bibitem{Sukhorukov}
A. A. Sukhorukov, Z. Xu, and Y. S. Kivshar,
{Phys. Rev. A} {\bf 82}, 043818 (2010).


\bibitem{Ablowitz1}
M.\ J.\ Ablowitz and Z.\ H.\ Musslimani, {Phys.\ Rev.\ Lett.}\ {\bf
110}, 064105 (2013).

\bibitem{Sarma}
A. K. Sarma, M. A. Miri, Z. H. Musslimani, and D. N.
Christodoulides,  {Phys. Rev. E}\  {\bf 89}, 052918 (2014).




\bibitem{Ablowitz2}
M. J. Ablowitz and Z. H. Musslimani, Phys. Rev. E {\bf 90}, 032912
(2014).

\bibitem{NLModels}
W. Liu, D. Q. Qiu, Z. W. Wu, and J. S. He, {Commun. Theor. Phys.} {\bf 65},  671 (2016); A. S. Fokas, {Nonlinearity} {\bf 29}, 319 (2016).

\bibitem{Khare2} A.\ Khare, A.\ Saxena, {arXiv: 1405.5267}.

\bibitem{LiXu}
M. Li and T. Xu, {Phys. Rev. E} {\bf 91}, 033202 (2015).

\bibitem{antidark} Yu. S. Kivshar, {Phys. Rev. A} {\bf43}, 1677 (1991).


\bibitem{Matveev1}
V.\ B.\ Matveev, M.\ A.\ Salle, {\em Darboux Transformations and
Solitons} (Springer Press, Berlin, 1991).

\bibitem{Matveev}
V. B. Matveev, {Phys. Lett. A} {\bf 166}, 205 (1992).


\bibitem{GDT}
B. L. Guo, L. M. Ling, and Q. P. Liu, {Phys. Rev. E} {\bf 85},
026607 (2012).

\bibitem{Roguewave}
R.\ Guo, Y. F. Liu, H. Q. Hao, and F. H. Qi, {Nonlinear Dyn.} {\bf 80}, 1221 (2015);
L. C. Zhao and J. Liu, {Phys. Rev. E}  {\bf 87}, 013201 (2013); L. Wang, C. Geng, L. L. Zhang, and Y. C. Zhao, {EPL} {\bf 108}, 50009 (2014); J. He, L. Guo, Y. Zhang, and A. Chabchoub, {Proc. R. Soc. London Series A} {\bf 470}, 20140318 (2014).


\end{thebibliography}
\end{document}